\documentclass[11pt,a4paper]{article}

\usepackage[utf8]{inputenc}
\usepackage[T1]{fontenc}
\usepackage{amsmath,amssymb,amsthm}
\usepackage{graphicx}
\usepackage{hyperref}
\usepackage{authblk}
\usepackage{enumitem}
\usepackage{dsfont}
\usepackage{mathtools}

\graphicspath{ {./images/} }

\usepackage{natbib}

\newcommand{\vect}[1]{\ensuremath{\boldsymbol{#1}}}
\def\plotmyimages{1}

\title{Straightforward Bayesian A/B testing with Dirichlet posteriors}

\author[1]{Dustin Hayden \thanks{Email: dustin.hayden@autotrader.co.uk}}
\author[1]{Tom Armitage \thanks{Email: tom.armitage@autotrader.co.uk}}

\affil[1]{Autotrader Research Group, Autotrader UK}

\date{\today}

\begin{document}

\maketitle

\begin{abstract}

Bayesian A/B testing investigates metric changes using the joint posterior distribution of two (or more) experimentally-derived datasets. The construction of said joint posterior is often a time-consuming process requiring specialized knowledge and domain expertise. In businesses that perform tens to hundreds of A/B tests per month it is important to have a robust analysis pipeline that can handle the variety of experiments performed on a modern web platform; requiring a domain expert to select appropriate prior and likelihood distributions for each experiment simply does not scale. In this work, we highlight a solution to this problem using a generalized approximation of the true joint posterior using a Dirichlet-Categorical model. While a manually-constructed, expert-tuned model for every dataset is preferable, the Dirichlet-Categorical approximation performs sufficiently well in both simulations and real-world scenarios to be internally used as the standard analysis method. 

\end{abstract}

\section{Introduction}

At Autotrader, teams of data scientists and business stakeholders work together to improve our website experience. Putative changes to the website are often run through a randomized A/B test to ensure the change performs well enough to put into production. Over time the number of experiments has grown, motivating improvements to the analysis process and, critically, to the communication of the results.

Several years ago, Autotrader switched to a Bayesian analysis of experimental tests for two main reasons: First, we wanted to improve the dialogue with our business stakeholders. When analyzing and presenting the results of the experiments, we found that stakeholder questions were more easily answered within a Bayesian framework. For instance, ``what is the probability Group A is better than Group B?''. We are quick to note that this is \textbf{not} what the $p$-value represents. There were also questions concerning risk-mitigation. For instance, ``if we're wrong and put into production the worse feature, what do we stand to lose?''. This question can be easily answered using a joint-posterior-derived value called expected loss (\cite{stucchio2015bayesian}), which we will further explain in Section \ref{subsec:loss}. 

Second, we wanted to encourage data scientists and stakeholders to flexibly adapt their pre-experimentally-determined statistical thresholds to an appropriate level of risk for a given metric or experiment. North Star metrics for big site changes should be risk-averse; curiosity metrics for small site changes can be more relaxed to focus on other business needs (e.g. more consistent branding, tackling tech debt). While this can be accomplished via changing the $p$-value threshold, such a change does not capture the severity of any error. There is also dogmatic reluctance to changing $p$-value thresholds rather than using the de facto $p < 0.05$.

Thus, we switched to a Bayesian, posterior-focused analysis method using the approach described in \cite{stucchio2015bayesian}, the outline of which is covered in \cite{atbayesblogpt1} and further elaborated throughout the course of this work. 

\begin{figure}[htbp]
  \centering
  \ifnum 1=\plotmyimages 
    \includegraphics[scale=1]{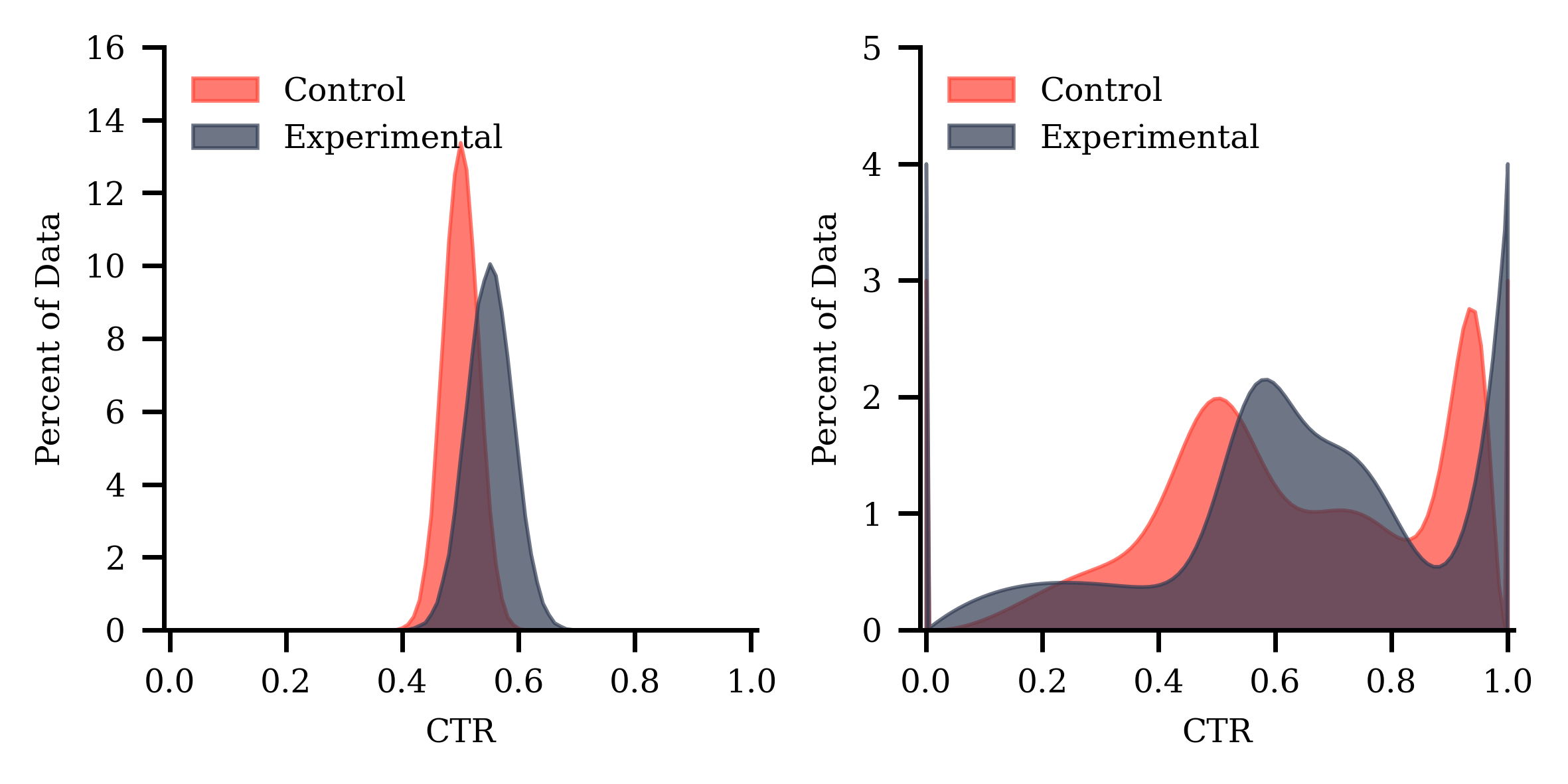}
  \else
    \includegraphics[width=12cm, height=6cm]{example-image-a}
  \fi
  \caption{Desirable, easily modelled data on the left and more realistic data on the right.}
  \label{fig:clean_v_real}
\end{figure}

In general, this Bayesian approach has worked well since its introduction at Autotrader. However, we noted several shortcomings that we wanted to address. The largest roadblock was the importance of selecting an appropriate conjugate prior and likelihood pair. For each experimental dataset, a data scientist must create a distribution that resembles the data (and ideally conduct exhaustive posterior predictive checks to ensure the model is representative). If the data behaves nicely, this is trivial. Further, if we assume each experimental group's data is independent, then the joint posterior becomes a more tractable product of individually-constructed conjugate pairs. Unfortunately, most data is not well represented by conjugate pairs, as illustrated in Fig. \ref{fig:clean_v_real}. In these plots, the click-through rate (CTR) in two scenarios is compared. The left subplot can use a simple posterior for both Control and Experimental data. The right subplot, showing more realistic user data, is less amenable to such a simple construction.

A viable solution needed to be quick, easily explainable, and extensible to many situations. One could construct a more complicated Hierarchical Bayesian model to describe the data and sample from these via a Markov chain Monte Carlo approach. However, constructing these models would take quite some time and requires some expertise in numerical statistics limiting how far such an approach could scale outside of the core group of specialists in the business. Similarly, Variational Inference (\cite{blei2017variational}) wasn't appealing given the need for correctly defining the family of distributions for approximation (not to mention the optimization algorithm needed to fit the candidate distributions). 

During this investigation period, some experiments were being analysed via a bootstrapping approach to get non-parametric estimates of metrics. The bootstrap itself can be thought of as a ``non-parametric, non-informative posterior distribution'' for a given parameter (\cite{hastie2017esl}, \cite{rubin1981bootstrap}). The idea of repeatedly sampling with replacement from the dataset was appealing. Even a small test at Autotrader can easily contain millions of logged events, so it is not objectionable to assume that this acquired data is fairly representative of the population. However, repeatedly sampling with replacement from millions of data points, not to mention the requirement for flat priors, caused a roadblock. Ideally a solution would reduce the memory and compute requirements and have the ability to include informative priors.

\begin{figure}[htbp]
  \centering
  \ifnum 1=\plotmyimages 
    \includegraphics[scale=1]{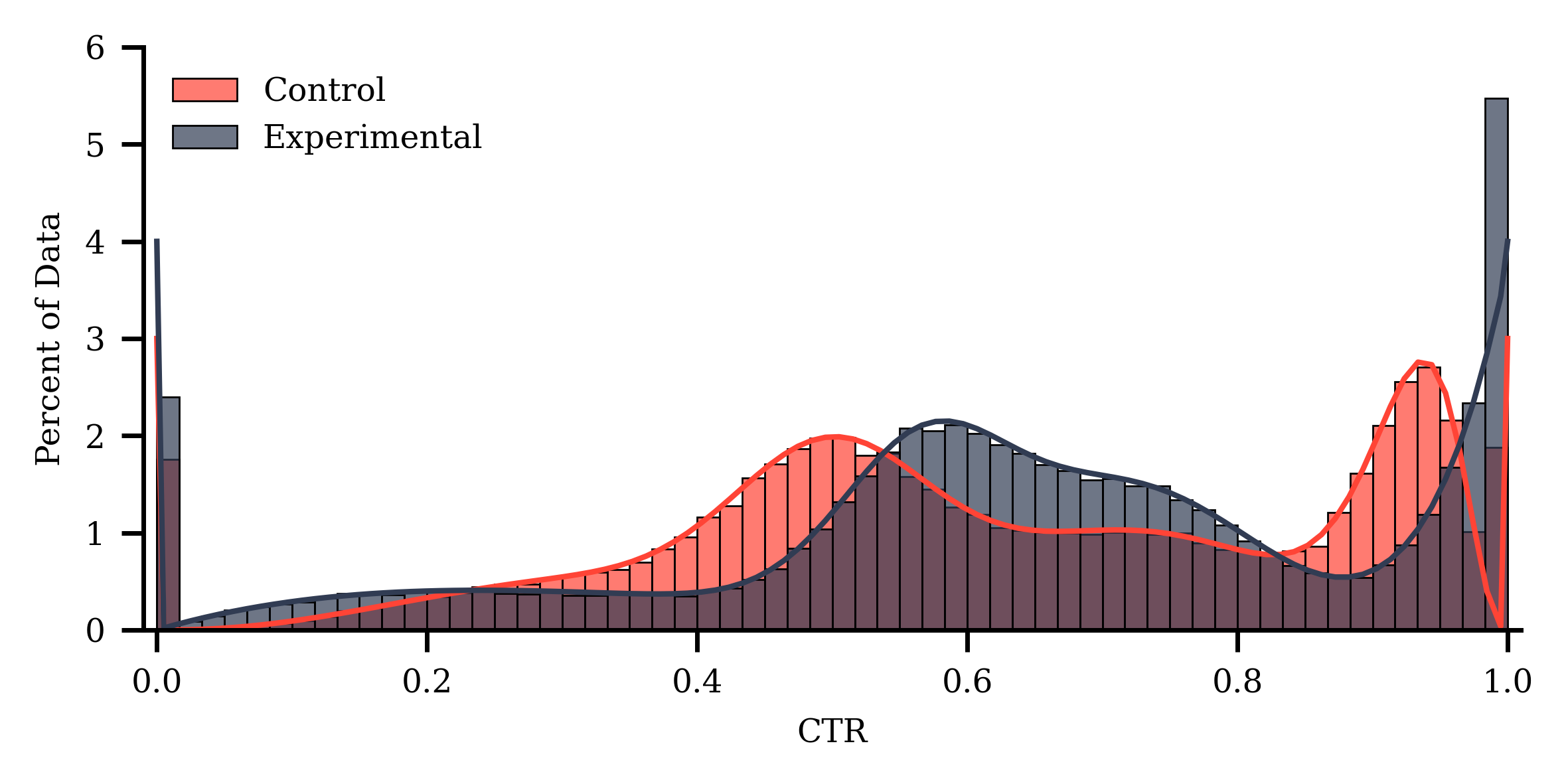}
  \else
    \includegraphics[width=12cm, height=6cm]{example-image-a}
  \fi
  \caption{Data can be binned to represent a Categorical distribution. The solid lines are the Riemannian-approximated probability density for the underlying tripartite hurdle model (details within Section \ref{sec:simulations}), normalized to have the same scale as the Categorical probability mass function.}
  \label{fig:real_binned}
\end{figure}

Our proposed solution was inspired by the treatment of revenue-based metrics in \cite{stucchio2015bayesian}, where multiple likelihoods are combined to better represent a real-world distribution. Specifically, these likelihoods separate the data into isolated chunks, originally accounting for an excess of users that performed no purchases at all. Could we generalize this approach and bin our raw observations into many discrete buckets? Most who analyse data are very familiar with histograms. Normalized histograms bin data into discrete bars whose height represents the proportion of the data within that bin such that all heights sum to 1, as illustrated in Fig. \ref{fig:real_binned}. This is a Categorical distribution in all but name. 

What if we chunk our observed data into bins, place it into a Categorical likelihood model, then use the conjugate Dirichlet prior to create a simple (but flexible) Dirichlet posterior? This so-called ``Bayesian histogram'' has been proposed as an easy, analytically tractable approximation of a given posterior (\cite{gelman2021bda}, Section 23.1). While we (and the authors of \cite{gelman2021bda}) are quick to note the limitations of such an approximation, we have found it quite useful at scale. It is simple to explain to less technical stakeholders (as most are familiar with histograms), simple to implement (common statistical programming libraries have functions for the Dirichlet distribution), and the posterior can quickly be sampled from to calculate various metrics.

This work details the Dirichlet-Categorical approximation for the joint posterior of a Bayesian A/B test. In Section \ref{sec:dirichlet-categorical}, we explain the Dirichlet distribution, how it relates to the Categorical distribution, and the Dirichlet-Categorical posterior (a.k.a. the ``Bayesian histogram''). In Section \ref{sec:metrics} we discuss a general form for calculating functions upon the Dirichlet-Categorical posterior as well as selected metrics like chance to beat, expected loss, and quantile differences. Finally, we test the quality of the approximations using simulations in Section \ref{sec:simulations}.

\section{The Dirichlet-Categorical Distribution}
\label{sec:dirichlet-categorical}

In this section we define the distributions used throughout the rest of the paper and how data can be processed using them. It is meant to be slow and introductory for data scientists not used to Dirichlet distributions. Those already comfortable with simplexes can likely skip to Section \ref{sec:metrics}.

\subsection{The Dirichlet Distribution}
\label{subsec:dirichlet}

\begin{figure}[htbp]
  \centering
  \ifnum 1=\plotmyimages 
    \includegraphics[scale=1]{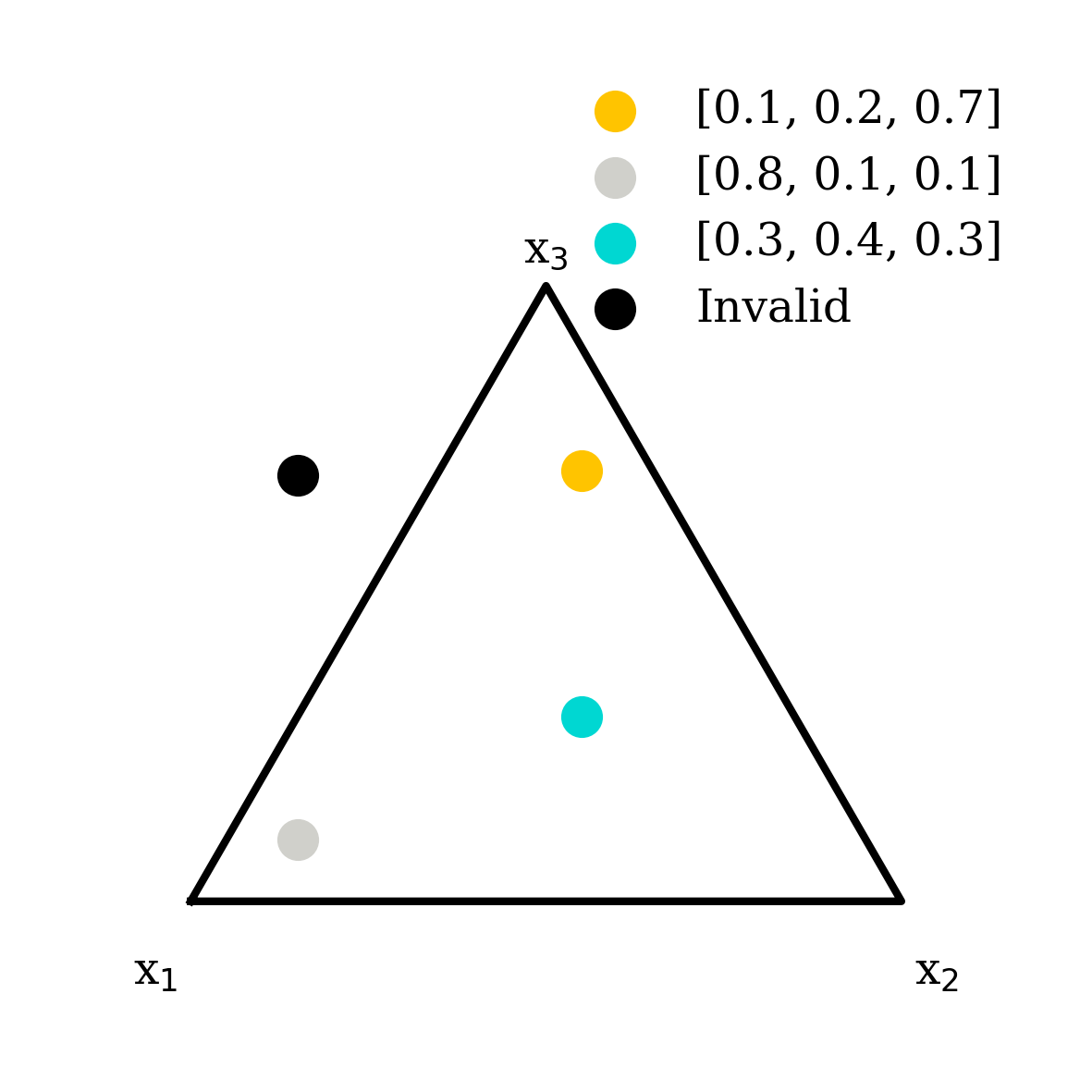}
  \else
    \includegraphics[width=6cm, height=6cm]{example-image-a}
  \fi
  \caption{Valid and invalid vectors on the $\Delta^{2}$-simplex.}
  \label{fig:simplex_points}
\end{figure}

Readers unfamiliar with Dirichlet distributions are likely to be confused by the following statement: a single draw from a random variable defined by a Dirichlet distribution returns a \textit{vector} of values. We will step through the distribution slowly to build intuition. A Dirichlet distribution is a probability density defined for $K\geq2$ elements in a vector. The support for a Dirichlet is the high-dimensional simplex $\Delta^{K-1}$:

\begin{equation}\label{simplex_rules}
    \Delta^{K-1} \equiv \left \{ 
    \vect{x} \in \mathbb{R}^{K} \colon 
    \sum_{i=1}^{K}x_{i}=1 \text{ and } x_{i} \in [0, 1] \textup{ for all } k 
    \right \}.
\end{equation}
In the syntax of machine learning, Eq.\eqref{simplex_rules} states that the valid input to a trained Dirichlet distribution model is a vector $\vect{x}$ whose elements sum to 1 wherein each element is bounded by 0 and 1. To better understand this, let's look at the situation where we have three elements in our vector $\vect{x}$. This creates a $\Delta^{2}$-simplex (also known as a triangle). Note that all the allowable vectors are constrained to this $\Delta^{2}$-simplex, as illustrated in Fig. \ref{fig:simplex_points}. 

The probability density for the Dirichlet distribution is defined by concentration parameters $\vect{\alpha}\ = (\alpha_{1}, ..., \alpha_{K})$ where $\alpha_{i} > 0$. The $\vect{\alpha}$ that describes a Dirichlet can be thought of as similar to the $\mu$ and $\sigma$ parameters that describe a Normal distribution (wherein you need both to describe the Normal distribution). Further, the $\vect{x}$ support for a Dirichlet is akin to the $x \in \mathbb{R}$ support for a Normal distribution (wherein there are only certain allowed inputs; you cannot evaluate the probability of a complex number using the standard Normal distribution). 

The density for the Dirichlet is given by:

\begin{equation}\label{dirichlet_pdf}
    f(\vect{x}; \vect{\alpha}) = 
    \frac{1}{B(\vect{\alpha})}
    \prod_{i=1}^{K}x_{i}^{\alpha_{i}-1},
\end{equation}
where $B(\vect{\alpha})$ is merely a normalizing constant: the multivariate beta function. Eq.\eqref{dirichlet_pdf} takes two equal-length vectors as input and returns a single probability density value. This single value represents the probability a random draw from the Dirichlet defined by $\vect{\alpha}$ yields $\vect{x}$.

As any given element, $\alpha_{i}$, in $\vect{\alpha}$ increases, it pulls the density towards the respective corner of the simplex (forcing a random draw to place more value on that $x_i$). To visualize this, let's again look at the $\Delta^{2}$-simplex. We plot the probability densities for the $\Delta^{2}$-simplex under multiple specifications of $\vect{\alpha}$. This entails defining a Dirichlet with a $\vect{\alpha}$, taking a swathe of values for the support $\vect{x}$, calculating the density using Eq.\eqref{dirichlet_pdf}, and plotting the resulting value at $\vect{x}$.

\begin{figure}[htbp]
  \centering
  \ifnum 1=\plotmyimages 
    \includegraphics[scale=1]{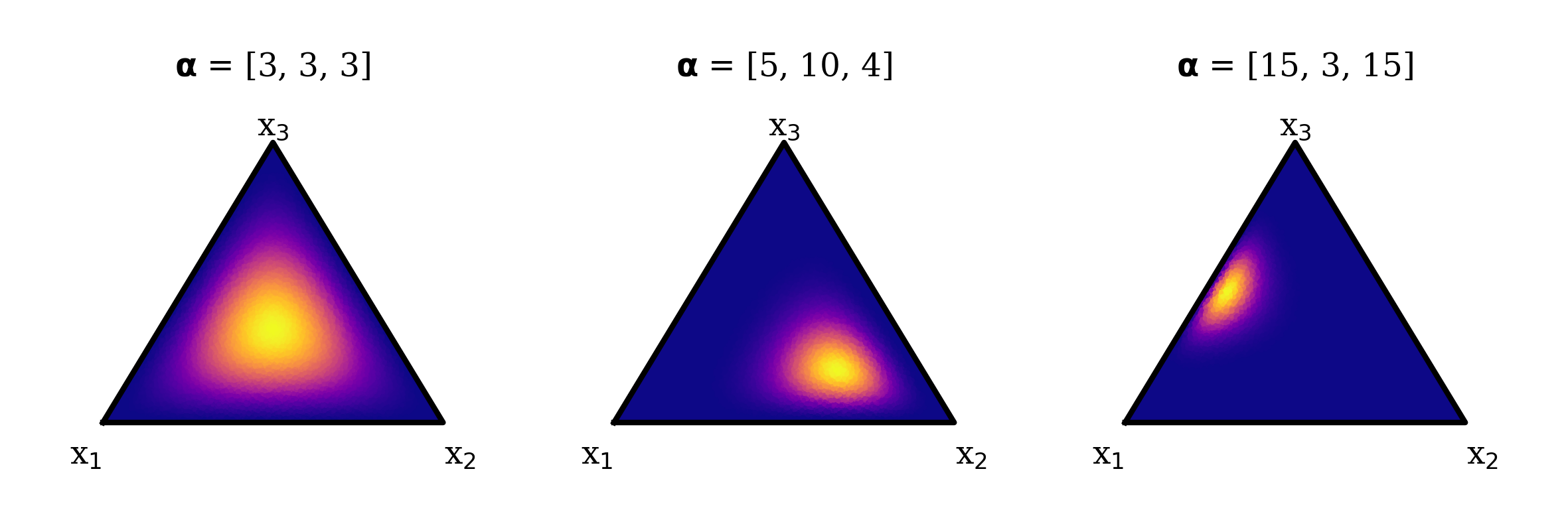}
  \else
    \includegraphics[width=12cm, height=4cm]{example-image-a}
  \fi
  \caption{Three examples of Dirichlet distributions. The quantitative colour scales are omitted, but hotter colours indicate more probability density assigned to the simplex location.}
  \label{fig:multiple_dirichlets}
\end{figure}

Fig. \ref{fig:multiple_dirichlets} demonstrates the probability of a given $\vect{x}$ being drawn from the Dirichlet. A multivariate random variable defined by drawing from a specified Dirichlet distribution is defined as $\vect{X} = (X_{1}, ..., X_{K}) \sim Dir(\vect{\alpha})$. The brighter the $\vect{x}$ location, the more probable the random variable $\vect{X}$ will take the value of that vector. This is often a source of confusion, so it's important to understand that $\vect{X}$ is a multivariate random variable that represents a location on the simplex $\Delta^{K-1}$. It is drawn with probability defined by $Dir(\vect{\alpha})$ (i.e. Eq.\eqref{dirichlet_pdf}). 

\begin{figure}[htbp]
  \centering
  \ifnum 1=\plotmyimages 
    \includegraphics[scale=1]{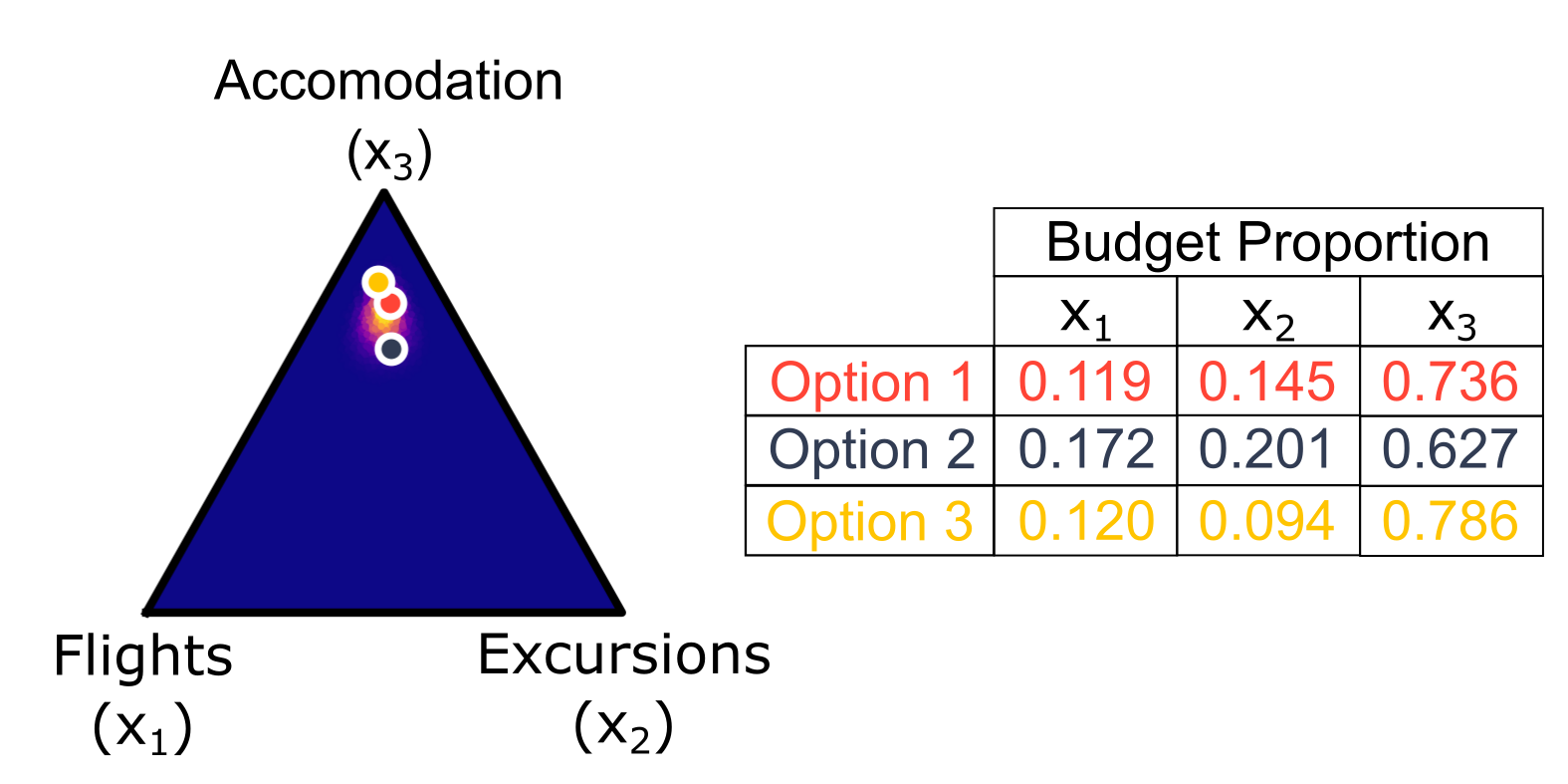}
  \else
    \includegraphics[width=8cm, height=4cm]{example-image-a}
  \fi
  \caption{Each random draw is the proportion of money spent.}
  \label{fig:vacation_example}
\end{figure}

As a way to explain to stakeholders: think of $\vect{X}$ as a constrained allocation of your budget for a holiday. You know you will spend exactly £1,000 in total on accommodation, flights and excursions. You'll likely spend most of the money on accommodation with the rest being evenly split between the flights and excursions. In other words, you have a proposed $Dir(\vect{\alpha})$ when you start planning, but you're uncertain what the exact amounts will end up being. As you look at cheaper accommodation you are able to allocate more towards flights and excursions. Each sample of your internal $Dir(\vect{\alpha})$ distribution generates a putative price-breakdown of your budget as seen in Fig. \ref{fig:vacation_example}.

The most important thing to take from this section is that random draws from (and inputs to) the Dirichlet will adhere to Eq.$\eqref{simplex_rules}$. Before we can understand how this is useful, however, we detour briefly to talk about the Categorical distribution.

\subsection{The Categorical Distribution}
\label{subsec:categorical}

The Categorical distribution is a discrete probability distribution for $K\geq1$ elements. It is defined by $\vect{p}\ = (p_{1}, ..., p_{K})$ where $p_{i} > 0$ and $\sum_{i=1}^{K}p_{i}=1$. Note that the stipulation for $\vect{p}$ is notably similar to what random draws from a Dirichlet would look like (Eq.$\eqref{simplex_rules}$). Each $p_{i}$ represents the probability that one of the mutually exclusive integers $z \in \{1, ..., K\}$ will be selected. The probability mass function can thus be written as:

\begin{equation}\label{categorical_pdf}
    f(z; \vect{p}) = 
    \prod_{i=1}^{K}p_{i}^{[z=i]},
\end{equation}
where $[z=i]$ is the Iverson bracket that evaluates to 1 when $z=i$ and 0 otherwise. 

\begin{figure}[htbp]
  \centering
  \ifnum 1=\plotmyimages 
    \includegraphics[scale=1]{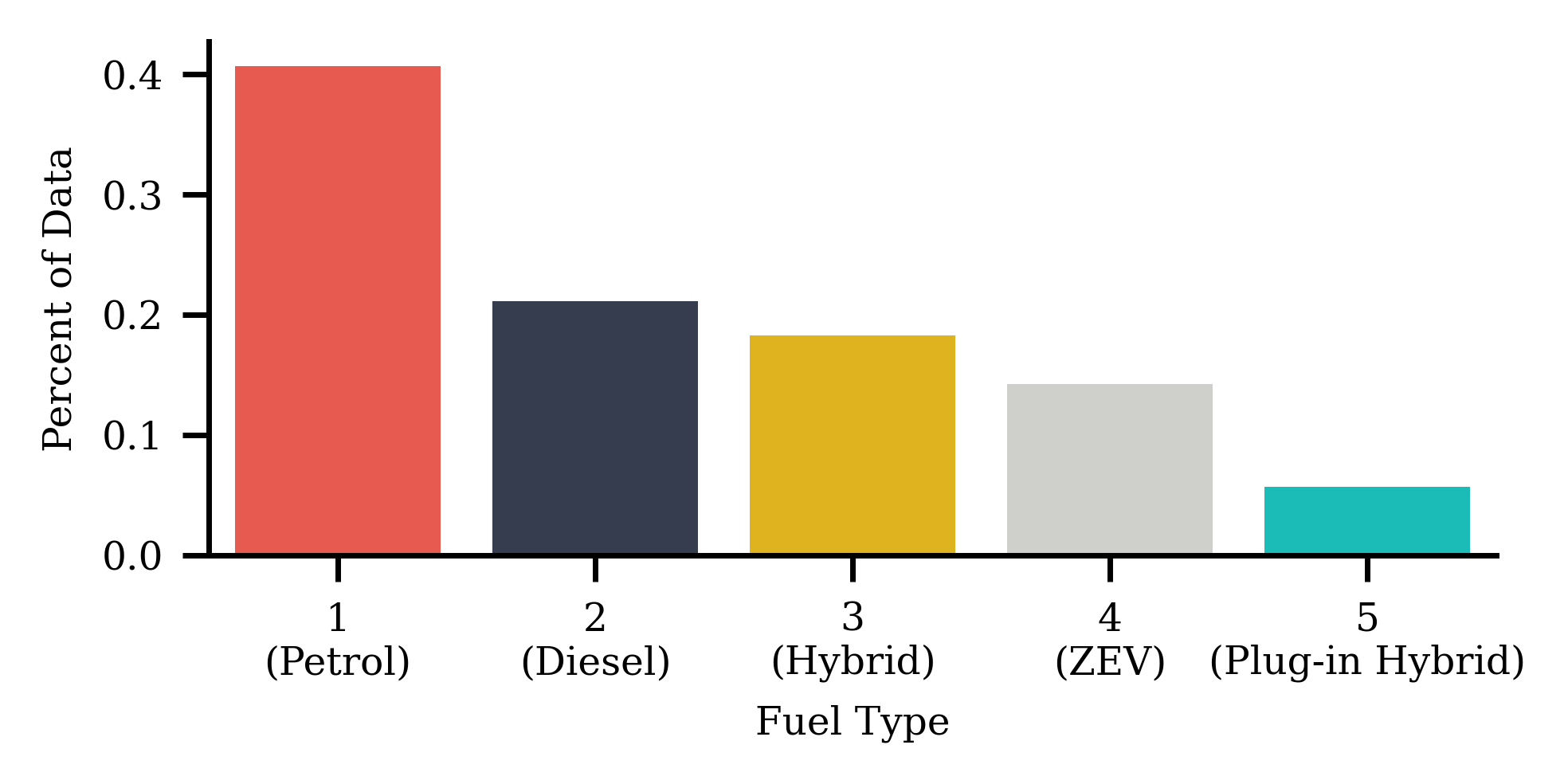}
  \else
    \includegraphics[width=10cm, height=5cm]{example-image-a}
  \fi
  \caption{Proportion of new UK vehicles registered with a fuel type in 2023 (Source: UK Government, Table VEH1153).}
  \label{fig:new_vehicle_categorical}
\end{figure}

A simple visual example is a normalized histogram of fuel types for sold automobiles in the UK in 2023 as seen in Fig. \ref{fig:new_vehicle_categorical}, where fuel type is represented as an integer. A random draw from a Categorical distribution is a single random integer (that might represent some other entity, like a fuel type). That is to say, $Z \in \{1, ..., K\} \sim Cat(\vect{p})$ where $\vect{p}$ must contain non-zero, positive numbers that sum to 1. 

\subsection{Dirichlet-Categorical Posterior}

In Section \ref{subsec:dirichlet} we discussed the fact a Dirichlet distribution will generate random number vectors that lie on the simplex. Section \ref{subsec:categorical} described how a Categorical distribution represents a normalized histogram of categories and that the required $\vect{p}$ is notably similar to the random variable vector $\vect{X}$ that is drawn from a Dirichlet.

It is thus unsurprising that the Dirichlet distribution is the conjugate prior distribution for the Categorical distribution. Imagine a set of data collected during an experiment which can be placed into one of $K$ mutually exclusive integers. The Dirichlet posterior that encapsulates a Dirichlet prior and a Categorical likelihood is:

\begin{equation}\label{dirichlet_posterior}
  \begin{aligned}
    \vect{\alpha} & = (\alpha_{1}, ..., \alpha_{K}) \text{ where } \alpha_{i} > 0 \\
    \vect{s} & = (s_{1}, ..., s_{K}) \text{ where } s_{i} = \sum_{i=1}^{K}[z=i] \\
    Dir(\vect{\alpha}+\vect{s}) & = Dir(\alpha_1+s_1, ..., \alpha_K+s_K),  
  \end{aligned}
\end{equation}
where $[z=i]$ is again the Iverson bracket that evaluates to 1 when $z=i$ and 0 otherwise. 

This breaks down a posterior into the element-wise sum of two vectors. The first vector $\vect{\alpha}$ encapsulates our prior thoughts and the second vector $\vect{s}$ is the count of our data within each category. If we had Categorical data, this would solve our requirement for a simple, easy to explain, and flexible posterior. We will work with this Categorical data assumption for a moment to elucidate the relationship between the Dirichlet and Categorical distributions.

A draw from the Dirichlet posterior is a random vector $\vect{X}$ that represents a Categorical distribution's vector parameter $\vect{p}$. Each time we draw from the Dirichlet, we create a new normalized histogram of Categorical data that represents the proportion of total data we would expect for each integer/category. 

\begin{figure}[htbp]
  \centering
  \ifnum 1=\plotmyimages 
    \includegraphics[scale=1]{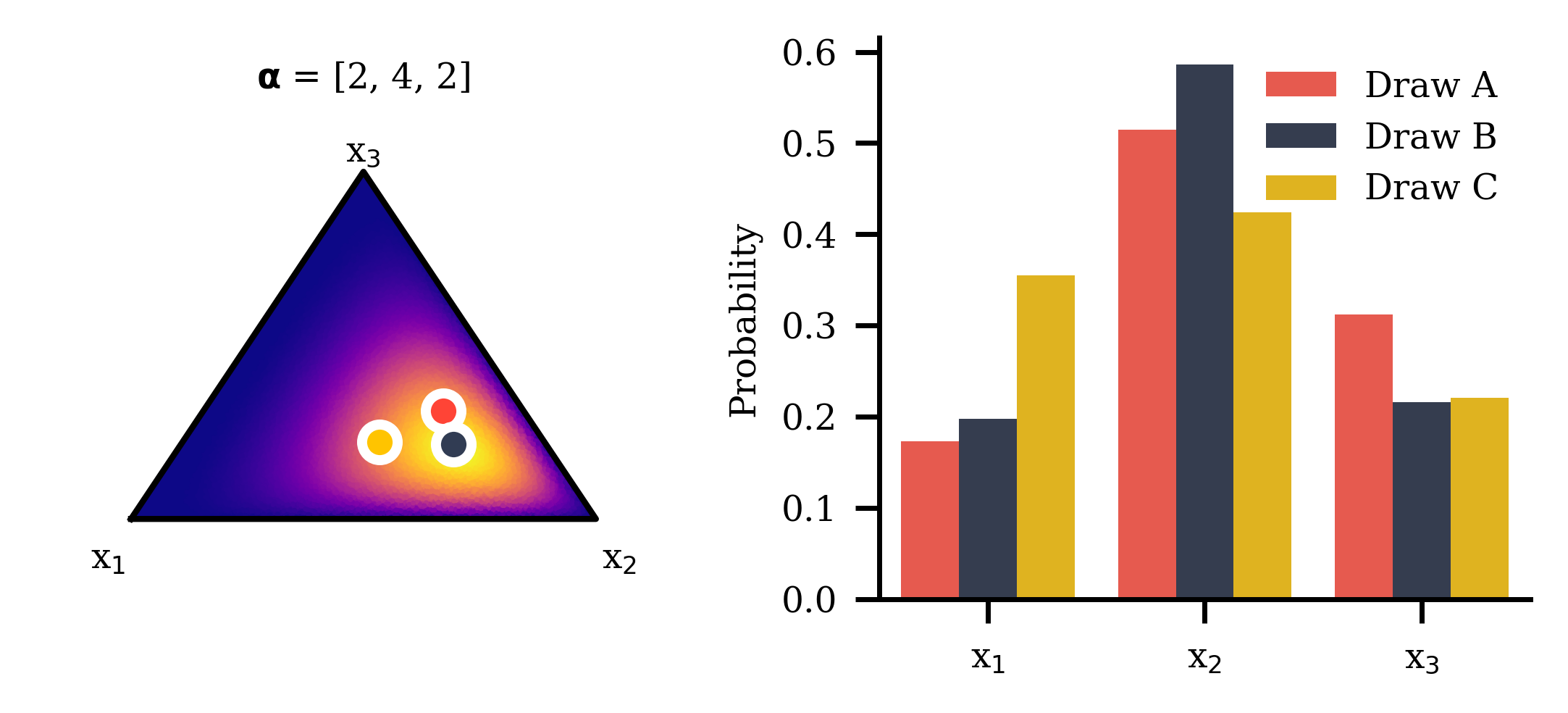}
  \else
    \includegraphics[width=11cm, height=5cm]{example-image-a}
  \fi
  \caption{Multiple random draws of a Dirichlet visualized on the simplex and a normalized histogram.}
  \label{fig:example_posterior_draws}
\end{figure}

We can return to the simple case of a $\Delta^{2}$-simplex for our intuition. There are three categories and we will use a flat prior $\vect{\alpha} = [1, 1, 1]$. For this example, let us assume 5 data points were obtained that yield $\vect{s} = [1, 3, 1]$. Using Eq.\eqref{dirichlet_posterior} we know our posterior is $Dir([2, 4, 2])$. We can construct such a Dirichlet, draw random samples from it, and visualize each sample as a normalized histogram (Fig. \ref{fig:example_posterior_draws}). Note how the normalized histograms all have more probability near the centre category than the edges (as does the data), but the exact shape and heights can vary (indicating the uncertainty in our estimate).

The more data we have (or the more informative our prior), the more tightly-concentrated our posterior. With increasing data the randomly drawn location vector $\vect{X}$ will be become quite stable and produce similar normalized histograms. Conversely, the less data there are, the more variable the random draw (and the resulting normalized histogram) will be.

But the data we presented in the introduction is not Categorical; it is a click-through rate. How does this Dirichlet-Categorical posterior framework help us in this case?

\subsection{Binning Data into the Dirichlet-Categorical Posterior}

In the introduction, we (and the authors of \cite{gelman2021bda}) noted that many of our visualizations of probability densities were based on normalized histograms created from sampling data (or some known posterior). We were taking continuous values/parameters and discretizing them for the purposes of graph-making. In other words, we were flipping our continuous variables into Categorical variables.

\cite{gelman2021bda} formalize this concept and we briefly review it here. Assume you have some bin boundaries $\vect{\xi} = (\xi_1, ..., \xi_{K+1})$ where $\xi_1 < \xi_2 < ... < \xi_K < \xi_{K+1}$. Note that the presence of one more boundary than categories ensures there will be $K$ resulting bins. Assume further that any given data $d_i$ you have is bounded by the extreme boundaries (i.e. $d_i \in [\xi_1, \xi_{K+1}]$). Under these assumptions, similar to Eq.\eqref{dirichlet_posterior}, we obtain

\begin{equation}\label{dirichlet_binned_posterior}
  \begin{aligned}
    \vect{\alpha} & = (\alpha_{1}, ..., \alpha_{K}) \text{ where } \alpha_{i} > 0 \\
    \vect{n} & = (n_{1}, ..., n_{K}) \text{ where } n_{i} = \sum_{j=1}^{N}\mathds{1}[\xi_i < d_j \leq \xi_{i+1}] \\
    Dir(\vect{\alpha}+\vect{n}) & = Dir(\alpha_1+n_1, ..., \alpha_K+n_K) \\
    Dir(\vect{\alpha^*}) & = Dir(\vect{\alpha}+\vect{n}),
  \end{aligned}
\end{equation}
where $\mathds{1}$ is the indicator function whose value is 1 when the inner bracket condition holds and 0 otherwise. The sum within the middle line is merely the number of observed data points that fall in the $i^{th}$ histogram bin.

\begin{figure}[htbp]
  \centering
  \ifnum 1=\plotmyimages 
    \includegraphics[scale=1]{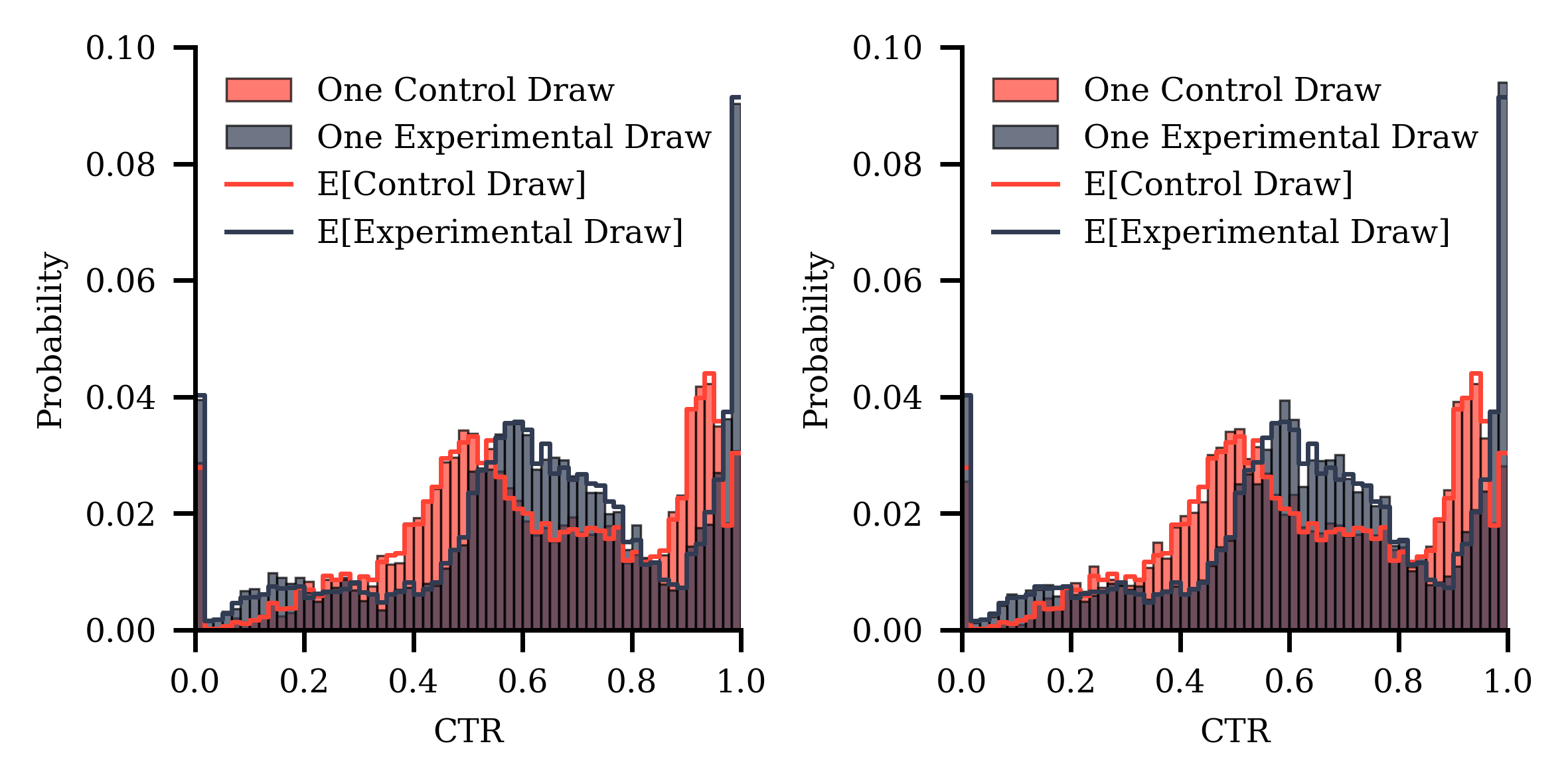}
  \else
    \includegraphics[width=12cm, height=6cm]{example-image-a}
  \fi
  \caption{Two sets of random draws from fit Dirichlet distributions visualized on separate normalized histograms. To help compare the draws, we also plot the expected values for each posterior as a line. The difference between the solid coloured lines and the histograms highlight the randomness of the draws.}
  \label{fig:real_posterior_draws}
\end{figure}

If we go back to the introductory example with click-through rates, we can use Eq.\eqref{dirichlet_binned_posterior} to construct a Dirichlet distribution of our data. We use $60$ bins for demonstration purposes. This creates a Dirichlet whose support is a $\Delta^{59}$-simplex, which we unfortunately have not found a way to visualize. However, we can show some of the normalized histograms that result from random draws of the posterior $Dir(\vect{\alpha^*})$. As you can appreciate in Fig. \ref{fig:real_posterior_draws}, for each draw, the histograms look similar, but differ slightly to indicate uncertainty.

We have approximated our more realistic data (which is a simulation of a hurdle model) into a posterior that is relatively simple to explain, create, and sample from. While a more advanced solution would be to use variational inference or hierarchical modelling to recreate each data set within an experiment, the practical demands of industry work preclude dedicating the requisite time needed to pursue those avenues for every A/B test conducted within a modern business.

Should a data scientist use Eq.\eqref{dirichlet_binned_posterior} they need only decide the location of the bin boundaries $\vect{\xi}$. At Autotrader, most of our analytical insights are necessarily bounded. Conversion rates, click rates, percent of total stock, etc. Due to the natural bounds being 0 and 1, and if we just use equally-spaced bin edges, the only decision becomes the number of bins to use. We dedicate Section \ref{sec:simulations} to exploring how the number of bins impacts the approximation. In short, the approximation works well for our metrics so long as you decide on a reasonable number of bins (i.e. more than 100 in realistic scenarios). In the case where metrics are unbounded the existence of a maximum bin value can aid with the handling of extreme values, such as those generated by bots.

Armed with an approximation of our posterior, let's analyse the joint posterior for an A/B test. While doing this we'll uncover one additional choice a data scientist must make: what ``value'' represents any given bin (i.e. the mean/median/mode of the data, the midpoint of the bin edges, etc).

\subsection{Choosing a Prior Vector}
Selecting an appropriate prior (i.e. selecting $\vect{\alpha}$ for Eq.\eqref{dirichlet_binned_posterior}) is beyond the scope of this white paper. The focus of this paper is to holistically and flexibly evaluate the full posterior distribution using a few key metrics that have been instrumental to experimentation at Autotrader. Thus, for all our reports and simulations, we use the uniform prior $\vect{\alpha} = K^{-1}\vect{1}$. We will briefly mention that, given our wealth of historical data, it is trivial to construct a prior of bin counts using a previous day's, week's, month's, or year's data. The more historical data used, the stronger the influence of the prior.

\section{Calculating Metrics}
\label{sec:metrics}
In Section \ref{sec:dirichlet-categorical} we created an approximation of the posterior for more realistic data. Now we use this posterior, rather multiple posteriors, to analyse the results of an A/B test. We present three metrics that have guided our decision making at Autotrader.

\subsection{General}

Making inferences about a probability distribution generally involves evaluating a function over the probability distribution. In our case, this is a multi-variable integral of the simplex. Given the (initial) complexity, we devote this section to a generalized process for evaluating a generic function upon the Dirichlet simplex. Luckily, we will finish the section with a simple-to-implement formula that can evaluate any function over the simplex via random sampling of the Dirichlet posterior. Common software packages make random sampling of Dirichlet distributions trivial, further reducing the complexity in a applied setting.

To begin, due to the high-dimensional space of a $\Delta^{K-1}$-simplex, we must introduce some notation. This will clean up the integration bounds making subsequent equations easier to interpret. As a first example, we will try to integrate all the probability associated with the simplex. Recall that the probability density function for a Dirichlet is given by Eq.\eqref{dirichlet_pdf} and by definition the integral must equal 1. Our Dirichlet posterior is defined by $\vect{\alpha^*}$ due to Eq.\eqref{dirichlet_binned_posterior}. Thus, we must integrate over the various values of the $\Delta^{K-1}$-simplex (i.e. all values the $K$-dimensional random vector $\vect{X}$ can take). This gives us:

\begin{equation}\label{simplex_integral_long_form}
    \int_{0}^{1}\int_{0}^{1-x_{1}} ... \int_{0}^{1-\sum_{i=1}^{K-1}x_{i}}
    f(\vect{x}; \vect{\alpha^*})
    dx_{K}...dx_{2}dx_{1}
    := 1.
\end{equation}
The bounded integrals of Eq.\eqref{simplex_integral_long_form} ensure that any vector $\vect{x} = [x_1, ..., x_K]$ that contributes to the inner function upholds Eq.\eqref{simplex_rules} (i.e. lie on the simplex $\Delta^{K-1}$). Starting from the leftmost integral, we state $x_1 \in [0, 1]$. $x_1$ must be $\geq 0$ to uphold the positivity and sum constraint of $Eq.\eqref{simplex_rules}$. The second leftmost integral is additionally constrained by our choice of $x_1$. To uphold the constraints of $Eq.\eqref{simplex_rules}$, $x_2 \in [0, [1-x_1]]$. By the inner integral, $x_K \in [0, [1-\sum_{i=1}^{K-1}x_{i}]]$, meaning $x_K$ is forced to take whatever positive value makes the entire vector sum to 1. Since this is just a feature of the simplex, we accumulate these selective integral bounds as $\int_{\Delta^{K-1}}...d\vect{x}$ and write:

\begin{equation}\label{simplex_integral_simple_form}
    \int_{\Delta^{K-1}}
    f(\vect{x}; \vect{\alpha^*})
    d\vect{x}
    := 1.
\end{equation}
Eq.\eqref{simplex_integral_long_form} and Eq.\eqref{simplex_integral_simple_form} are equivalent. Both state that, if we integrate the probability function Eq.\eqref{dirichlet_pdf} for the Dirichlet defined by $\vect{\alpha^*}$ over the simplex, the result by definition is 1. Eq.\eqref{simplex_integral_simple_form} provides a nice shorthand upon which to evaluate various functions of the posterior, forming the basis of our Bayesian inference. 

Recall that a multidimensional random variable drawn from a $\Delta^{K-1}$-simplex is $\vect{X} = (X_{1}, ..., X_{K}) \sim Dir(\vect{\alpha^*})$. In our setup, the random variable $\vect{X}$ represents the relative proportions of data that would fall into a given bin of a Bayesian histogram. If we wish to determine the effect of some function $g(\cdot)$ upon the \textit{proportions} vector $\vect{X}$, we evaluate the result as follows:

\begin{equation}\label{simplex_integral_general_form}
    \mathbb{E}[g(\vect{X})] = \int_{\Delta^{K-1}}
    g(\vect{x})
    f(\vect{x}; \vect{\alpha^*})
    d\vect{x}.
\end{equation}
While it might look daunting, this is just the definition for expected values acting on functions of random variables. Intuitively, the expected value of any function acting upon the simplex is the result of the function at a given simplex location weighted by how likely it is said location will be selected. 

We quickly note that functions upon $\vect{X}$ are merely functions of the proportions of data, not the values represented by those proportions. Before we address this, there is one extra bit of notation we must introduce. Bayesian inference in the context of A/B testing requires two posteriors: one for Group A and one for Group B. Luckily, since we can safely assume both sets of data are independent of each other (network effects that cause test buckets to interact with each other are not common at Autotrader), this is a simple matter of adding subscripts,

\begin{equation}\label{AB_simplex_integral_general_form}
    \begin{split}
        \mathbb{E}[g(\vect{X}_A, \vect{X}_B)] = \\
        \int_{\Delta_A^{K_A-1}} \int_{\Delta_B^{K_B-1}}
        g(\vect{x}_A, \vect{x}_B)
        f(\vect{x}_A; \vect{\alpha_A^*})
        f(\vect{x}_B; \vect{\alpha_B^*})
        d\vect{x}_B
        d\vect{x}_A,
    \end{split}
\end{equation}
while a bit complex in notation, the above is relatively straight-forward. In English, Eq.\eqref{AB_simplex_integral_general_form} integrates (twice) over all possible values of the simplexes - once for Group A and once for Group B. It allows for each simplex to have its own number of categories (though in practice we always use the same number for both). It then gets the joint probability (assuming independence) for the specific pair $(\vect{x}_A, \vect{x}_B)$ using their independently-created Dirichlet posteriors. Finally, it applies some function $g(\cdot)$ to the $(\vect{x}_A, \vect{x}_B)$ proportions pair.

Analytically solving any function with Eq.\eqref{AB_simplex_integral_general_form} is guaranteed to be difficult if not intractable due to the integrals. One option would be to discretize the simplex and estimate the integral as a Riemannian sum. Another would be to use a sampling approach. We opt for the latter, given the ubiquity and ease of Dirichlet sampling within modern statistical software packages. 

In the sampling appraoch, we first generate $N$ randomly sampled vectors $\vect{x}$ from each posterior. Group A is drawn from $\vect{X}_{A} \sim Dir(\vect{\alpha_A^*})$, giving us $\vect{x}_{A, j}$ for $j \in {1,...,N}$. Group B similarly creates $\vect{x}_{B, j}$ for $j \in {1,...,N}$ by drawing from $\vect{X}_{B} \sim Dir(\vect{\alpha_B^*})$. We pair these $N$ samples, apply some function $g(\cdot)$ on them, and get the mean value across all samples. Mathematically this is given by

\begin{equation}\label{AB_simplex_integral_approx_general_form}
    \mathbb{E}[g(\vect{X}_A, \vect{X}_B)] 
    \approx 
    \frac{1}{N}\sum_{j=1}^{N}g(\vect{x}_{A, j}, \vect{x}_{B, j}).
\end{equation}
An added benefit of the simulation approach of Eq.\eqref{AB_simplex_integral_approx_general_form} is that you can plot the result of $g(\cdot)$ upon each random draw, naturally visualizing the variation in the estimated expected value. 

There is one final bit of bookkeeping we need to attend to. As we stated earlier, both Eq.\eqref{AB_simplex_integral_general_form} and Eq.\eqref{AB_simplex_integral_approx_general_form} do not include information about what each category/integer/bin represents. They are merely vectors representing the proportion of data that will fall into said category/integer/bin. What we need is a dictionary mapping of index to value. This can be unique for each Dirichlet posterior, should you choose. With a slight abuse of notation, we state such a mapping is $v(\vect{X}) \in \mathbb{R}^K$.

\begin{figure}[htbp]
  \centering
  \ifnum 1=\plotmyimages 
    \includegraphics[scale=1]{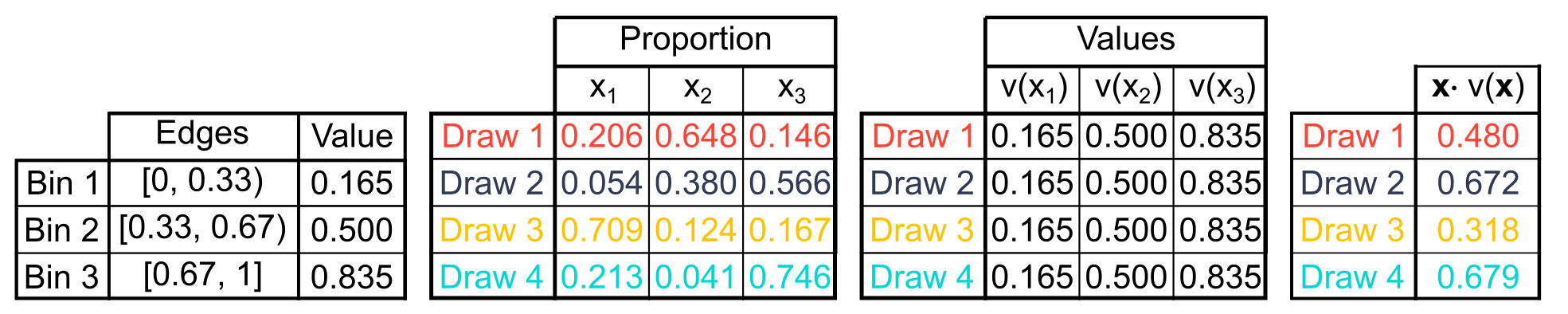}
  \else
    \includegraphics[width=10cm, height=2cm]{example-image-a}
  \fi
  \caption{A demonstration of the mapping process. The rightmost table is the dot product of the vector of proportions and the value vector.}
  \label{fig:mapping_example}
\end{figure}

For example, using 3 equally-spaced bins, $X_1$ is the first element of $\vect{X}$, representing the proportion of the data that would end up within the interval $[0, 0.33]$. It does $\mathbf{not}$ represent the \textit{value} of said bin, merely how much of the data is likely to end up there. To get the values for $\vect{X}$, we look it up in the dictionary $v(\vect{X})$, as illustrated in Fig. \ref{fig:mapping_example}. As an example, we'll use the midpoint of a bin's interval as the mapping value.

The mapping dictionary $v(\vect{X})$ is determined by the data scientist. It can be the left edge, right edge, or midpoint of the bin. Or it can be some statistic of the data within the bin, such as the mean, median, or mode. In practice we use the median value of all data within the bin or the midpoint of the bin's range if there is no data within the bin. In practice, we use enough bins such that the determination of the mapping is nigh inconsequential. Whatever the mapping is, it causes the equations to change slightly. Namely, instead of applying a function $g(\cdot)$ upon a realization of the random variable for proportions $\vect{X}$, we apply the function to the proportions $\vect{X}$ and the value mapping $v(\vect{X})$. We call this a weighted value mapping function, $w(\cdot)$:

\begin{equation}\label{weighted_mapping_function}
    w(\vect{x}_A, \vect{x}_B) 
    = g(\vect{x}_A, \vect{x}_B, v_A(\vect{x}_A), v_B(\vect{x}_B)).
\end{equation}

The above just improves our shorthand and emphasizes that any generic function $g(\cdot)$ we are attempting to evaluate over the simplex must account for \textbf{both} the proportions $\vect{X}$ as well as the values $v(\vect{X})$. Adding this slightly changes our generic formulae.

We first have:

\begin{equation}\label{AB_simplex_integral_general_form_with_values}
    \begin{split}
        \mathbb{E}[w(\vect{X}_A, \vect{X}_B)] = \\
        \int_{\Delta_A^{K_A-1}} \int_{\Delta_B^{K_B-1}}
        w(\vect{x}_A, \vect{x}_B)
        f(\vect{x}_A; \vect{\alpha_A^*})
        f(\vect{x}_B; \vect{\alpha_B^*})
        d\vect{x}_B
        d\vect{x}_A
        ,
    \end{split}
\end{equation}
which we can approximate with:

\begin{equation}\label{AB_simplex_integral_approx_general_form_with_values}
    \mathbb{E}[w(\vect{X}_A, \vect{X}_B)]
    \approx 
    \frac{1}{N}\sum_{j=1}^{N}w(\vect{x}_{A,j}, \vect{x}_{B,j}).
\end{equation}
The notation (and journey to get to Eq.\eqref{AB_simplex_integral_approx_general_form_with_values}) belies the simplicity of the system. In essence, we can conduct (approximate) Bayesian inference using three steps. First, get a Dirichlet posterior for each experimental group using the straightforward equation Eq.\eqref{dirichlet_binned_posterior}. Second, create a function $w(\cdot)$ that accounts for both the proportion of data within each bin and the values representing each bin as in Eq.\eqref{weighted_mapping_function}. Finally, draw random samples from each group's Dirichlet posterior and use Eq.\eqref{AB_simplex_integral_approx_general_form_with_values} to estimate the expected value of the function.

We hope that, once the notation is understood and we provide a few examples, you can appreciate the simple beauty of such an approximation framework. We (and others) are quick to note the limitations and caveats of such a system. Namely, it is best used on bounded metrics where a minimum and maximum are both known and guaranteed. Further, the values within a bin may not be uniformly distributed across the bin's interval, making the choice of $v(X_i)$ a source of bias. This bias can be mitigated with a larger number of bins, wherein the intervals are so narrow that the choice of $v(X_i)$ is nigh-inconsequential. Additionally, this technique is unlikely to work well when data is sparse or rare. However, in the context of our analytical work at a technology company like Autotrader, we have found the above system to work remarkably well, be easily understood, and was rapidly rolled out and adopted by the entire analytics department.

As an interesting aside, this is a general form of the Bayesian bootstrap (\cite{hastie2017esl}, \cite{rubin1981bootstrap}). To conduct a Bayesian bootstrap, use a flat prior, take each data point to be its own bin (even if it shares a value with other data points), and use the data point's value as the mapping. In such a case, this procedure can be thought of as drawing weights to be applied to each sample (i.e. the Bayesian bootstrap).

\subsection{Chance to Beat (Population Average)}
\label{subsec:chancetobeat}

Probably the most common question we receive from business stakeholders is ``what is the chance our change made things better''? A simple first usage of Eq.\eqref{AB_simplex_integral_approx_general_form_with_values} addresses this question. 

The stakeholders, at face value, want to know $Pr(Experimental > Control)$. However, often when data scientists and stakeholders work with probability distributions, this is simplified to merely a comparison of the expected values for each group. In other words, while asked about $Pr(E > C)$, people instinctively switch the question to $Pr(\overline{E} > \overline{C})$. We continue this time-honoured tradition as an edifying first example.

There are two groups in our click-through rate example from the introduction: Control and Experimental. We decide to create 100 equally spaced bins and we decide that the mapping from bin-to-value will be the median value of the data within the bin. We use Eq.\eqref{dirichlet_binned_posterior} to get $\vect{\alpha_C^*}$ using the Control data and $\vect{\alpha_E^*}$ using the Experimental data. We then create our maps $v_C(\cdot)$ and $v_E(\cdot)$.

Now we must decide how to represent a function that captures $Pr(\overline{E} > \overline{C})$. We first need a function that finds the weighted mean of the value mapping $v(\vect{x})$ where the weights are $\vect{x}$. Technically this is what $\overline{\vect{x}}$ represents, but its routine use as the sample mean requires us to use separate notation to make our intentions clear:
\begin{equation}\label{overline_equation}
    \widetilde{\vect{x}} = \sum_{i=1}^{K}x_{i}v(x_i),
\end{equation}
where $x_i$ represents the proportion of data in the bin and $v(x_i)$ represents the value of that bin. 

\begin{figure}[htbp]
  \centering
  \ifnum 1=\plotmyimages 
    \includegraphics[scale=1]{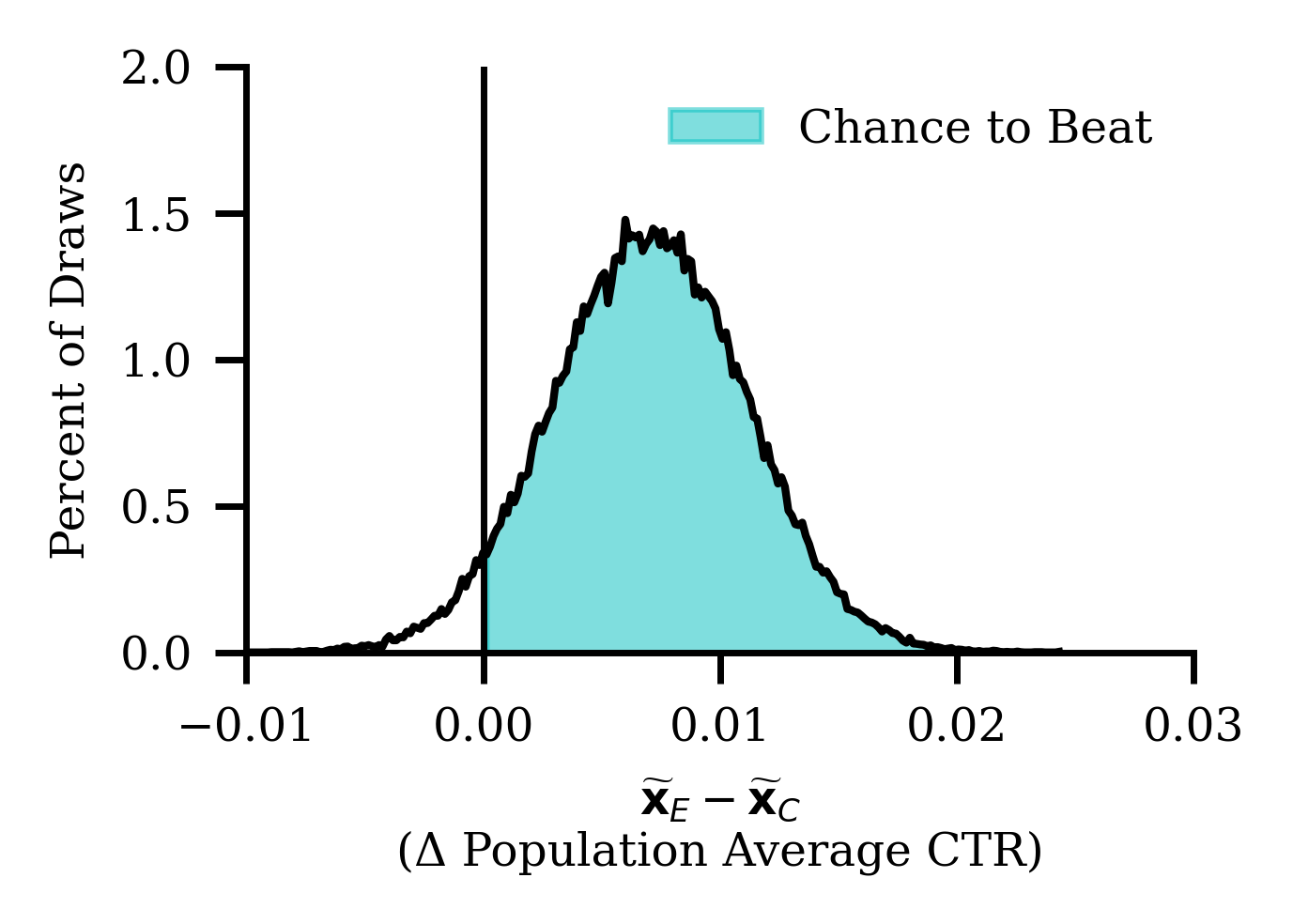}
  \else
    \includegraphics[width=7cm, height=5cm]{example-image-a}
  \fi
  \caption{Sample draws for the chance to beat of the average.}
  \label{fig:chance_to_beat_draws}
\end{figure}

Using the notation in Eq.\eqref{overline_equation} and an indicator function we can create a $w(\cdot)$ that captures whether any given random draw of vectors has a better average:

\begin{equation}\label{average_chance_to_beat_internal_function}
    w_{\overline{E}>\overline{C}}(\vect{x}_E, \vect{x}_C) = \\
    \mathds{1}[\widetilde{\vect{x}}_{E} > \widetilde{\vect{x}}_{C}].
\end{equation}
This function will return 1 when the mean of the experimental group is higher than the control group and 0 otherwise. If we evaluate that function over the simplex, we obtain our chance to beat measure. We can get an exact answer by plugging Eq.\eqref{average_chance_to_beat_internal_function} into Eq.\eqref{AB_simplex_integral_general_form_with_values}:

\begin{equation}\label{average_chance_to_beat_exact}
    \begin{split}
        Pr(\overline{E} > \overline{C}) = \\
        \mathbb{E}[w_{\overline{E}>\overline{C}}(\vect{X}_E, \vect{X}_C)] = \\
        \int_{\Delta_E^{99}} \int_{\Delta_C^{99}}
        \mathds{1}[\widetilde{\vect{x}}_{E} > \widetilde{\vect{x}}_{C}]
        f(\vect{x}_E; \vect{\alpha_E^*})
        f(\vect{x}_C; \vect{\alpha_C^*})
        d\vect{x}_C
        d\vect{x}_E
        ,
    \end{split}
\end{equation}
which we can approximate using Eq.\eqref{AB_simplex_integral_approx_general_form_with_values}:
\begin{equation}\label{average_chance_to_beat_approx}
    Pr(\overline{E} > \overline{C}) \approx 
    \frac{1}{N}\sum_{j=1}^{N}
    \mathds{1}[\widetilde{\vect{x}}_{E, j} > \widetilde{\vect{x}}_{C, j}].
\end{equation}
Visually, we can calculate the difference in the averages for $100,000$ draws, as in Fig. \ref{fig:chance_to_beat_draws}. The chance to beat (Eq.\eqref{average_chance_to_beat_approx}) is the total normalized area under the curve to the right of $\widetilde{\vect{x}}_{E} - \widetilde{\vect{x}}_{C} = 0$.

While the chance to beat is a nice metric, it has some issues. Imagine a scenario in which our distributions for the means are wider, as in Fig. \ref{fig:chance_to_beat_flaw}. We perform the same estimation process and plot all the random draws from the trained Dirichlet posteriors:

\begin{figure}[htbp]
  \centering
  \ifnum 1=\plotmyimages 
    \includegraphics[scale=1]{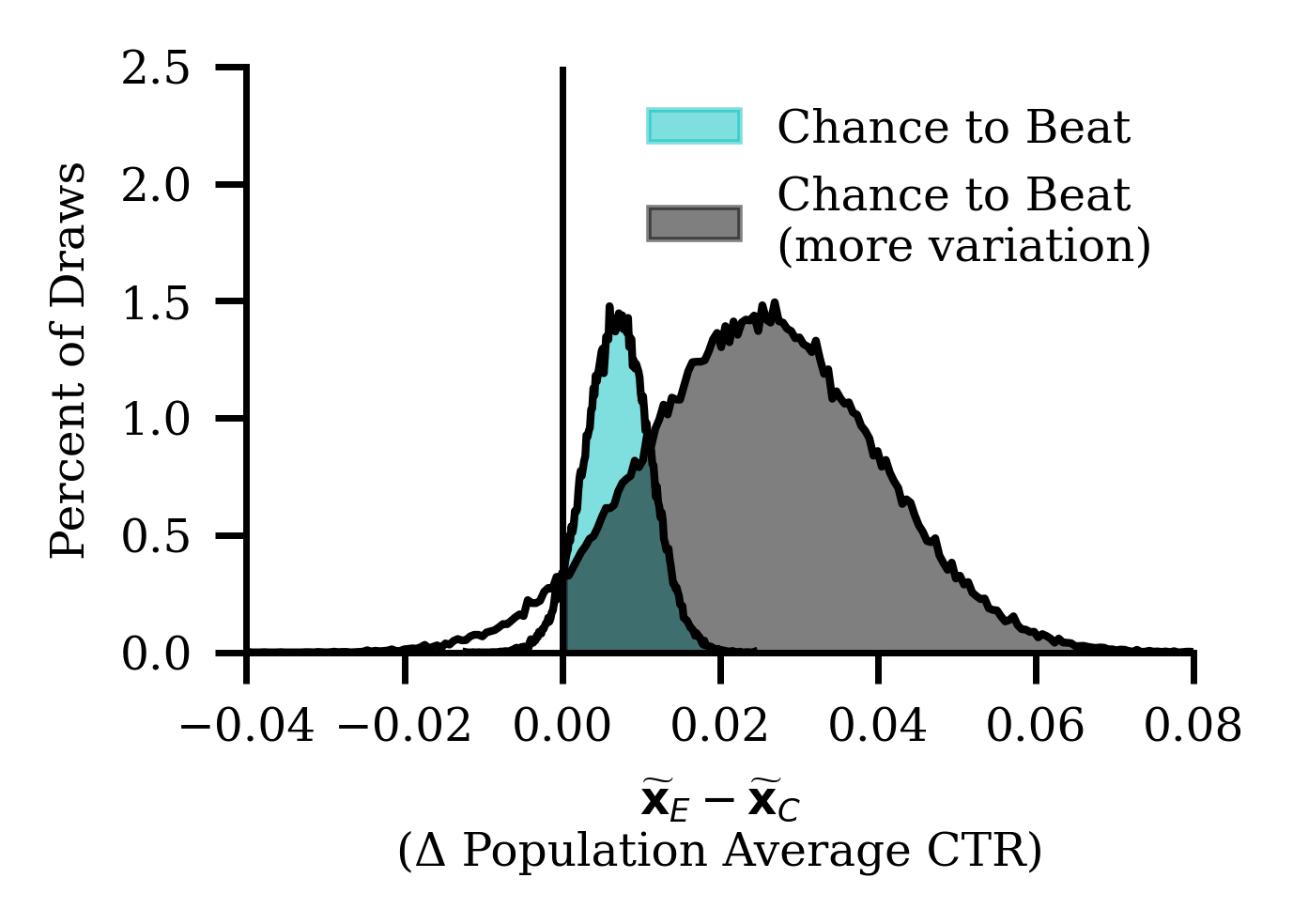}
  \else
    \includegraphics[width=7cm, height=5cm]{example-image-a}
  \fi
  \caption{Chance to beat fails to account for the magnitude of the difference.}
  \label{fig:chance_to_beat_flaw}
\end{figure}

Both of the distributions within Fig. \ref{fig:chance_to_beat_flaw} have the same chance to beat (i.e. normalized area under the curve to the right of $\widetilde{\vect{x}}_{E} - \widetilde{\vect{x}}_{C} = 0$). In fact for both it's $95\%$ of the total area. However, the wider distribution represents a high-risk-high-reward scenario. The expected value of the difference in the wider distribution is higher, but it co-occurs with increased probabilities towards more extreme negative differences (i.e. the Control version being much better than the Experimental version). In other words, if you happen to be wrong in the wider-variation version, you stand to lose a lot more than the narrow-variation version. Simply reporting the $Pr(\overline{E} > \overline{C})$ might not be a sound business decision. 

So how do we account for this potential risk? One obvious option is to calculate and scrutinize the variance of the mean. In our experience, such a process is often treated as a nuisance and doesn't quite focus on the issue. The issue is that we're ignoring the \textit{magnitude} of the difference. Take another look at Eq.\eqref{average_chance_to_beat_internal_function} and note that it's a binary decision. Either it's bigger or it isn't. 
Is there some way to capture the magnitude of any difference as well as how likely it is that said difference will be observed?

\subsection{Expected Loss (Population Average)}
\label{subsec:loss}

We highly recommend the technical white paper by \cite{stucchio2015bayesian} that motivated our switch to a Bayesian form of analysis and inference. The author defines \textit{expected loss}, a metric that captures both the chance of a metric being a certain value and the severity of the impact if the metric was said value. As expected from the name, the expected loss specifically focuses in on the curve where $\widetilde{\vect{x}}_{E} - \widetilde{\vect{x}}_{C} < 0$ (i.e. when the Control version is actually better than the Experimental version).

\begin{figure}[htbp]
  \centering
  \ifnum 1=\plotmyimages 
    \includegraphics[scale=1]{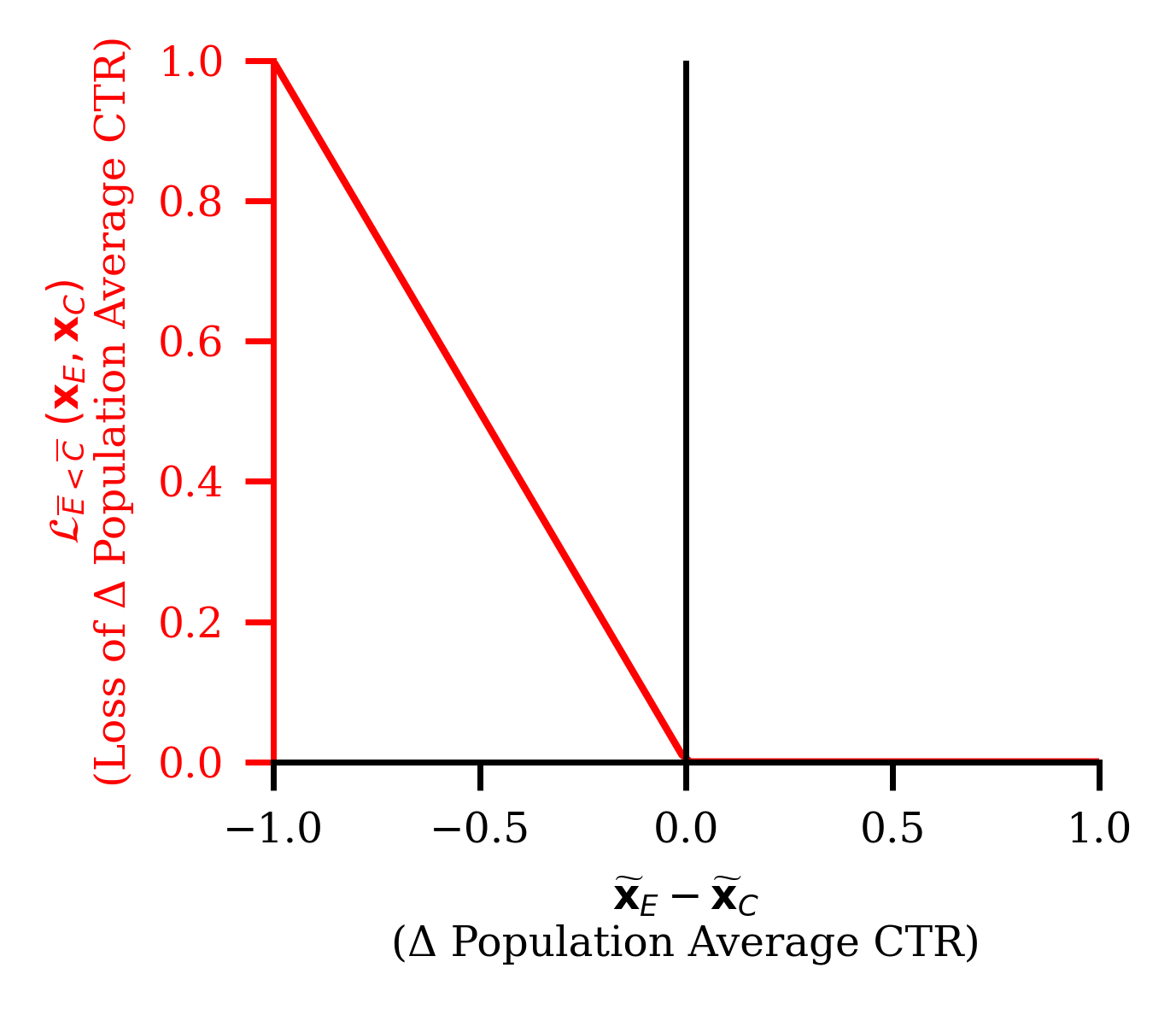}
  \else
    \includegraphics[width=7cm, height=6cm]{example-image-a}
  \fi
  \caption{A generic visualization of the reflected ReLU loss function (for the loss of choosing Experimental over Control).}
  \label{fig:loss_function}
\end{figure}

Since we're talking about loss, and to keep with the notation of (\cite{stucchio2015bayesian}), we'll switch from defining $w(\cdot)$ to $\mathcal{L}(\cdot)$. In our notation, the loss functions for the means are:

\begin{equation}\label{stucchio_loss_EC}
    \mathcal{L}_{\overline{E}<\overline{C}}(\vect{x}_E, \vect{x}_C) = \\
    \mathds{1}[\widetilde{\vect{x}}_{E} < \widetilde{\vect{x}}_{C}]
    (\widetilde{\vect{x}}_{C} - \widetilde{\vect{x}}_{E}),
\end{equation}

\begin{equation}\label{stucchio_loss_CE}
    \mathcal{L}_{\overline{C}<\overline{E}}(\vect{x}_E, \vect{x}_C) = \\
    \mathds{1}[\widetilde{\vect{x}}_{C} < \widetilde{\vect{x}}_{E}]
    (\widetilde{\vect{x}}_{E} - \widetilde{\vect{x}}_{C}).
\end{equation}
The above equations consider two situations. The first is the loss if the Experimental group's mean is worse than the Control group's mean, the second is vice versa. In either case, the indicator function earmarks only situations in which the appropriate group is worse and the simple difference function returns the severity of the loss. We visualize $\mathcal{L}_{\overline{E}<\overline{C}}(\vect{x}_E, \vect{x}_C)$ in Fig. \ref{fig:loss_function}.

Readers familiar with machine learning will recognize Fig. \ref{fig:loss_function} as a reflected rectified linear unit acting upon the value $\widetilde{\vect{x}}_{E} - \widetilde{\vect{x}}_{C}$. When the Experimental group's weighted mean is better than the Control group's, there is no loss. However, when it is lower, the loss is the absolute difference between the means. This makes loss a value that is always positive and in the units of the data generated, allowing for easy interpretation and intuition. The loss can be made negative if desired, especially when plotting. 

Continuing with our example from Section \ref{subsec:chancetobeat} where we are using 100 bins and have two Dirichlet posteriors. We can get an exact answer for the expected loss of choosing the Experimental group over the Control group by plugging Eq.\eqref{stucchio_loss_EC} into Eq.\eqref{AB_simplex_integral_general_form_with_values}:

\begin{equation}\label{average_expected_loss_exact}
    \begin{split}
        \textit{Expected Loss (E)} = \\
        \mathbb{E}[\mathcal{L}_{\overline{E}<\overline{C}}(\vect{X}_E, \vect{X}_C)] = \\
        \int_{\Delta_E^{99}} \int_{\Delta_C^{99}}
        \mathds{1}[\widetilde{\vect{x}}_{E} < \widetilde{\vect{x}}_{C}]
        (\widetilde{\vect{x}}_{C} - \widetilde{\vect{x}}_{E})
        f(\vect{x}_E; \vect{\alpha_E^*})
        f(\vect{x}_C; \vect{\alpha_C^*})
        d\vect{x}_C
        d\vect{x}_E
        ,
    \end{split}
\end{equation}
which we can approximate using Eq.\eqref{AB_simplex_integral_approx_general_form_with_values}:

\begin{equation}\label{average_expected_loss_approx}
    \textit{Expected Loss (E)} \approx 
    \frac{1}{N}\sum_{j=1}^{N}
    \mathds{1}[\widetilde{\vect{x}}_{E,j} < \widetilde{\vect{x}}_{C,j}]
    (\widetilde{\vect{x}}_{C,j} - \widetilde{\vect{x}}_{E,j}).
\end{equation}
An equivalent process can create:

\begin{equation}\label{average_expected_loss_approx_alt}
    \textit{Expected Loss (C)} \approx 
    \frac{1}{N}\sum_{j=1}^{N}
    \mathds{1}[\widetilde{\vect{x}}_{C,j} < \widetilde{\vect{x}}_{E,j}]
    (\widetilde{\vect{x}}_{E,j} - \widetilde{\vect{x}}_{C,j}).
\end{equation}
As noted previously, the values received from Eq.\eqref{average_expected_loss_approx} and Eq.\eqref{average_expected_loss_approx_alt} are in the same units as the data being modelled. A 0 indicates the entire (sampled) probability space for the Experimental group's mean is above the Control group's mean. The exact value of an estimate is how much your mean will, on average, drop should you choose to go with the Experimental (or Control) treatment and the Experimental (or Control) happened to be worse. 

\begin{figure}[htbp]
  \centering
  \ifnum 1=\plotmyimages 
    \includegraphics[scale=1]{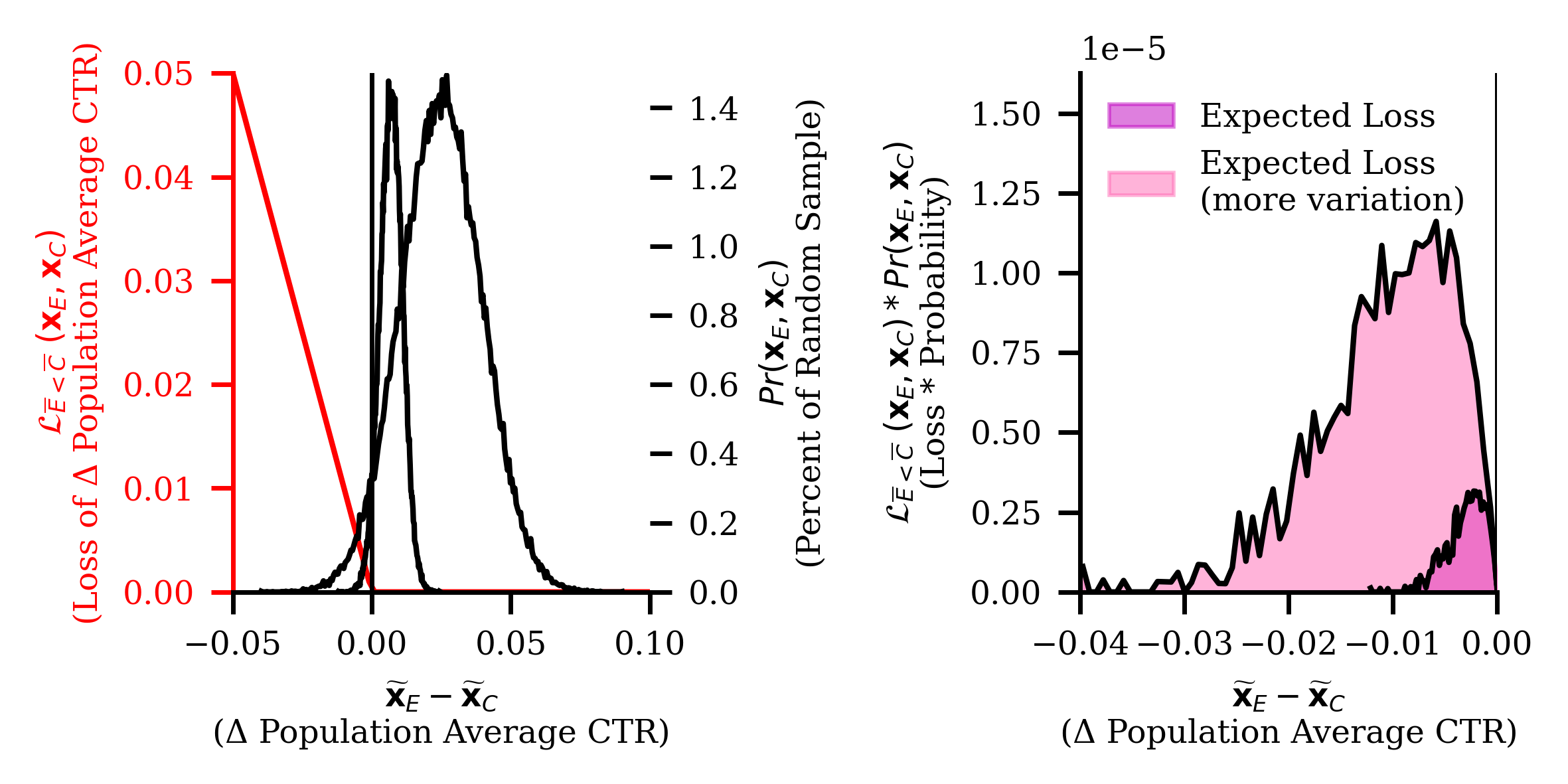}
  \else
    \includegraphics[width=12cm, height=6cm]{example-image-a}
  \fi
  \caption{While the chance to beat for the averages were equal, expected loss highlights the increased risk and variation of the wider distribution.}
  \label{fig:expected_loss}
\end{figure}

We can return to our illustrative example from Section \ref{subsec:chancetobeat}. We once more plot the sampled range of averages for each group. However, we also apply Eq.\eqref{stucchio_loss_EC} to each sample and plot the resulting values. As you can see in Fig. \ref{fig:expected_loss}, the loss is sensitive to both the probability and the magnitude, highlighting risky situations. 

The language of expected loss is naturally suited to business contexts. Often there are other business needs or processes that support a switch to the Experimental group (more consistent branding, relieving tech debt, etc). Instead of asking ``is the Experimental group clearly better'' and only deploying once you have evidence to discount the null case based on an arbitrary threshold, you can instead ask ``would the switch be detrimental to our metrics'' and adjust your risk levels as necessary. This allows for rapid development iteration where distinctive wins are hard to attain and a lack of change in the data is allowable (or in some cases even preferred).

\subsection{Quantiles}
\label{subsec:quantiles}

It does not escape our attention that functions upon the probabilities of population averages could easily be estimated using the central limit theorem. Indeed, we ourselves exploit this in our simulations as a method of comparison. 

But what if we want to look beyond the mean? Focusing on only the averages can hide re-distributions of probabilities. These re-distributions fuel hypotheses; perhaps a change has improved a metric for one subset of visitors while decreasing it for another subset? Perhaps the overall average has increased but at the expense of more low-CTR visitors (being balanced out by more high-CTR visitors)? Can we use that information to tailor our site towards each group specifically, improving the experience for all?

To assess distribution changes, the quantile function can be used. The quantile function at a given value $\tau$ returns the infimum of the cumulative distribution function (CDF) for a random variable. In other words, it finds the smallest value for the random variable such that the cumulative probability up until said value is $\tau$. This is represented as: 

\begin{equation}\label{quantile_equation}
    Q(\tau) = inf\{x \in \mathbb{R}: \tau \leq CDF(x)\}.
\end{equation}
Most are already familiar with the value for $Q(0.5)$; it is the median of a distribution for a random variable.

If we had a single random variable to represent our data, for instance if we modelled it as being drawn from a Gaussian, then we could directly use Eq.\eqref{quantile_equation}. However, care must be taken in the Dirichlet-Categorical approximation. We assume data is being generated from some true unknown (or computationally intractable) distribution. In the approximation framework, we estimate this distribution using a random variable vector $\vect{X}$ of proportions and a mapping function of values for each proportion in the vector $v(\vect{X})$. Critically, this means we are not interested in the CDF of the Dirichlet distribution, which is preferable since to our knowledge there is no closed form solution (or stable approximation method) of a Dirichlet's CDF. Rather, we are interested in the cumulative proportions within any given sample. This is a subtle and important distinction. 

Any given draw of $\vect{X}$ sums to 1. We can thus use a draw, $\vect{x}$, within the quantile function as a proxy for the true CDF of the distribution we're approximating. The objective is to find the lowest possible index into the cumulative sum of sample $\vect{x}$ such that it is greater than the desired $\tau$. We can then use our value mapping dictionary $v(\vect{x})$ to extract the value associated with the indexed bin. Note that we implicitly assume (and enforce during usage) that the bins are ordered low to high in the Dirichlet-Categorical framework. This sample quantile is defined to be:

\begin{equation}\label{quantile_sample_equation}
    Q_{\vect{x}}(\tau) \coloneqq v(x_i) : 
    inf\{i: \tau \leq \sum_{j=1}^{i}x_j\}.
\end{equation}
The uncertainty around the quantile, or the difference between the Experimental group's and Control group's quantile, can be approximated using the same Dirichlet framework as the chance to beat and expected loss. For one final time, we'll walk through the process.

First, we need to create our weighted value mapping function (Eq.\eqref{weighted_mapping_function}) that utilizes information from $\vect{x}_C$, $\vect{x}_E$, $v_C(\vect{x}_C)$, and $v_E(\vect{x}_E)$. This will simply be the difference between the sample quantile for the experimental draw and the sample quantile for the control draw:

\begin{equation}\label{delta_sample_quantile}
    \Delta Q_{\vect{x}_E, \vect{x}_C}(\tau) = 
    Q_{\vect{x}_E}(\tau) - Q_{\vect{x}_C}(\tau).
\end{equation}

Next, we place this into the simplex integral (Eq.\eqref{AB_simplex_integral_general_form_with_values}) to get an exact answer for the expected value of Eq. \eqref{delta_sample_quantile}:

\begin{equation}\label{delta_quantile_exact}
    \begin{split}
        \mathbb{E}[\Delta Q_{\vect{X}_E, \vect{X}_C}(\tau)] = \\
        \int_{\Delta_E^{99}} \int_{\Delta_C^{99}}
        \Delta Q_{\vect{x}_E, \vect{x}_C}(\tau)
        f(\vect{x}_E; \vect{\alpha_E^*})
        f(\vect{x}_C; \vect{\alpha_C^*})
        d\vect{x}_C
        d\vect{x}_E
        ,
    \end{split}
\end{equation}

which can be approximated using Eq.\eqref{AB_simplex_integral_approx_general_form_with_values}:

\begin{equation}\label{delta_quantile_approx}
    \mathbb{E}[\Delta Q_{\vect{X}_E, \vect{X}_C}(\tau)] \approx 
    \frac{1}{N}\sum_{j=1}^{N}
    \Delta Q_{\vect{x}_{E, j}, \vect{x}_{C, j}}(\tau).
\end{equation}

\begin{figure}[htbp]
  \centering
  \ifnum 1=\plotmyimages 
    \includegraphics[scale=1]{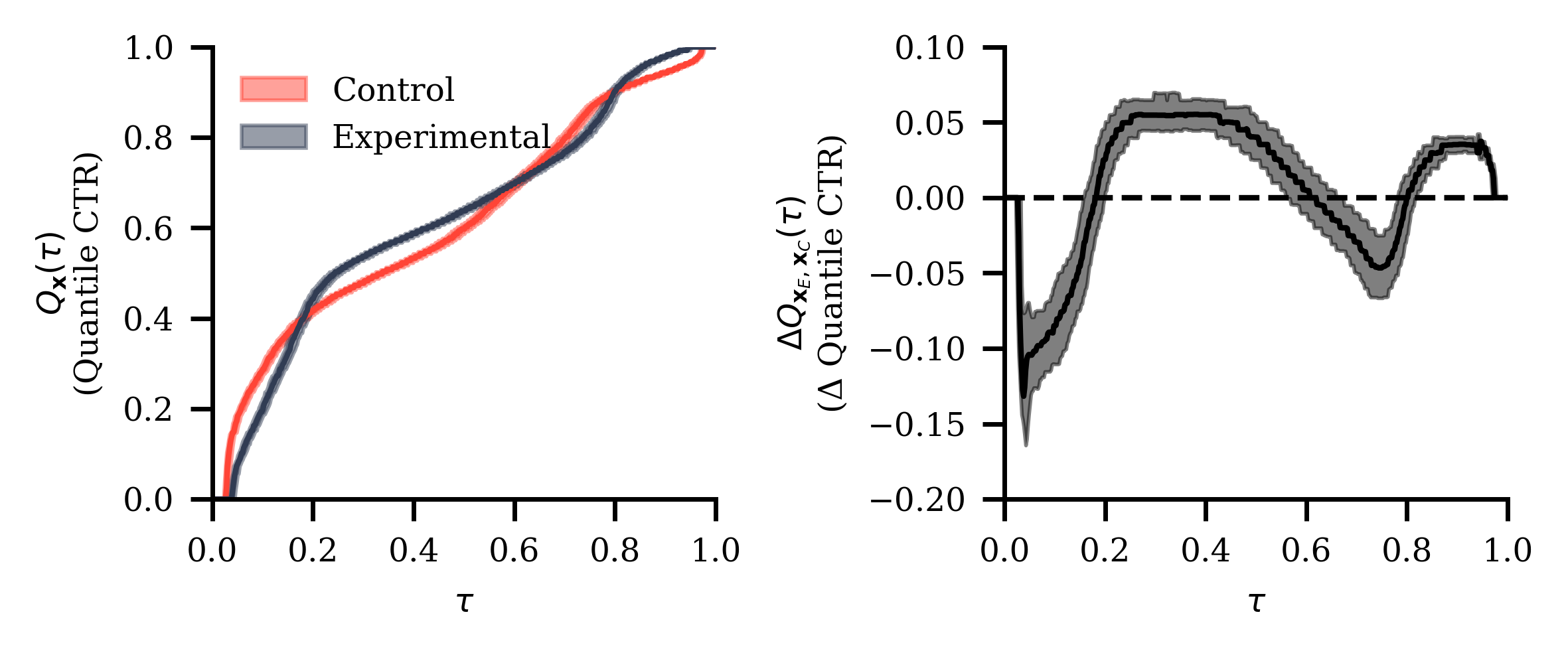}
  \else
    \includegraphics[width=12cm, height=5cm]{example-image-a}
  \fi
  \caption{Quantile plots reveal the different dynamics of the probability distributions. Shaded regions represent the $99\%$ credibility interval.}
  \label{fig:quantile_differences}
\end{figure}

In Fig. \ref{fig:quantile_differences}, we plot the quantiles for the CTR of both Experimental and Control data, as well as their difference. The right graph reveals multiple key insights that the average does not. First, the Experimental group has more visitors with a 0-CTR, as evidenced by the initial sharp reduction in the quantile difference plot near low $\tau$ values (where $Q_{\vect{x}_{E}(\tau)} = 0$ while $Q_{\vect{x}_{C}(\tau)} > 0$). Second, the difference climbs until $\tau = 0.2$ where the sign flips, indicating that up until the $20^{th}$ percentile the Control data has a higher CTR. Third, the Experimental overtakes until roughly the the $60^{th}$ percentile, meaning that the median value for CTR is higher in the Experimental group. Fourth, after the slight negative dip, there is another increase in the difference, indicating that the Experimental group has more visitors with higher CTRs.

Altogether, this means that the Experimental group is on average, or rather on median, better than the Control group ($\Delta Q_{\vect{x}_E, \vect{x}_C}(0.5)>0$), but with the caveat that there is an increase in 0-CTR and low-CTR visitors ($\Delta Q_{\vect{x}_E, \vect{x}_C}(\tau<0.2)<0$) that are being balanced out by an increase in high-CTR visitors ($\Delta Q_{\vect{x}_E, \vect{x}_C}(\tau>0.8)>0$). Such a result can prompt data scientists at Autotrader to segment visitors and seek out reasons for the bifurcated response. It might even lead to unique website changes tailored to certain segments of visitors to improve their experience without sacrificing another visitor segment's CTR.

We appreciate the journey to this point has been long and the notation dense, but we hope this final metric has emphasized the adaptability and relative simplicity of the Dirichlet approximation procedure. At Autotrader, the chance to beat, expected loss, and quantile plots all contribute to our understanding of experimental results and play an integral role in our increasingly rapid experimentation pipeline.

\section{Simulations}
\label{sec:simulations}

To provide guidance on the number of bins to use and to test the reliability of the framework, we ran simulations where a true (messy) distribution was able to be created and assessed with an alternative strategy as a reference-point. In this section, we detail these simulations. We encourage readers interested in adopting this methodology to run their own simulations on data resembling theirs before switching.

\subsection{Methods}
A major difficulty with verification is the creation of ground truth values to compare the Dirichlet-Categorical approximations to. By design, this framework tackles difficult-to-model posteriors whose ground truth values are similarly difficult to determine. We considered massive sampling of hierarchical Bayesian models using MCMC methods, but the computational costs precluded more than a handful of fine-tuned simulations. Instead, we settled on a tripartite hurdle model involving a zero-mass, a beta mixture model with $B$ betas, and a one-mass. For each simulation, two independent hurdle models are generated. Each hurdle model is defined by simulation parameters $p_{0_s}$, $\vect{\theta}_s$, and $p_{1_s}$ and is formulated as:

\begin{equation}\label{hurdle_model}
    Pr(X_s = x | p_{0_s}, \vect{\theta}_s, p_{1_s}) = \left\{ \begin{array}{lcr}
    p_{0_s} & \mbox{for} & x=0 \\ 
    f(x|\vect{\theta}_s) & \mbox{for} & 0 < x < 1 \\
    p_{1_s} & \mbox{for} & x=1
    \end{array}\right.
\end{equation}
where

\begin{equation}\label{beta_mixture_function}
f(x|\vect{\theta}_s) = \sum_{j=1}^{B}p_{s,j}Beta(x; \alpha_{s,j}, \beta_{s,j})
\end{equation}

and $\sum_{j=1}^{B}p_{s,j} = 1 - (p_{0_s} + p_{1_s})$.

For each simulation, the number of visitors (and thus samples from each hurdle model) was a random integer between $8,000$ and $25,000$. This represents a single day of $1\%$ of website traffic we commonly use for initial A/B tests. Besides the number of visitors, each hurdle model's parameters were randomly and independently generated. First, the hurdle probabilities $p_{0_s}$, $\sum_{j=1}^{B}p_{s,j}$, and $p_{1_s}$ were randomly sampled from a Dirichlet defined by $\vect{\alpha} = [1, 1, 1]$. The number of beta distributions within the mixture model was uniformly sampled from the integers $1$ through $15$. The proportion of probability assigned to each $p_{s,j}$ within $\sum_{j=1}^{B}p_{s,j}$ was randomly sampled from a Dirichlet defined by a vector of ones of length $B$. Each beta distribution's $\alpha_{s,j}$ and $\beta_{s,j}$ were randomly sampled values between $1.01$ and $100$ to ensure the support for the resulting Beta mixture model is $x \in (0, 1)$ (i.e. $p_{0_s}$ and $p_{1_s}$ are the only contributors to the probability for 0 and 1). 

These parameters were chosen to qualitatively yield a broad swathe of varying distributions similar to the ones Autotrader routinely sees. The ranges were chosen before seeing the results, but we acknowledge we may have luckily stumbled into a behaving subset of parameter space.

Samples representing visitors were generated from the hurdle model using the inverse transform sampling method with a Riemannian-based estimation of the PDF and CDF using 1 million points evenly spread on the interval $[0, 1]$. To obtain the closest ground truth values we could, we used these Riemannian chunks to calculate the chance to beat, expected loss, and quantile differences for the two randomly-generated hurdle distributions representing each group. These Riemannian-estimated numbers are referenced as either the ``True Value''s or the ground truth in subsequent sections. The Dirichlet-Categorical framework was used on the generated samples from these distributions and compared to the ground truth for the various statistics. We used the median value of all data in the bin (or the midpoint if no data was in the bin) as our choice of $v(\vect{X})$. As mentioned before, with a small number of bins, this can potentially bias results due to non-uniform data within a given bin interval. Given our findings below, we do not believe this is an appreciable source of bias in our datasets if enough bins are used. 

To benchmark the performance of the Dirichlet-Categorical approximation, we utilised two separate methods. For the first method, we invoked the central limit theorem and calculated the Normal distribution for the population average difference (using the sample means and the unbiased sample variance). This was used for the chance to beat and expected loss comparisons. For the second method, we bootstrapped the population average difference and the quantile differences. These were used for the chance to beat, expected loss, and quantile comparisons.

We ran $10,000$ simulations and present the results in the following section.

\subsection{Results}
\begin{figure}[!htbp]
  \centering
  \ifnum 1=\plotmyimages 
    \includegraphics[scale=1]{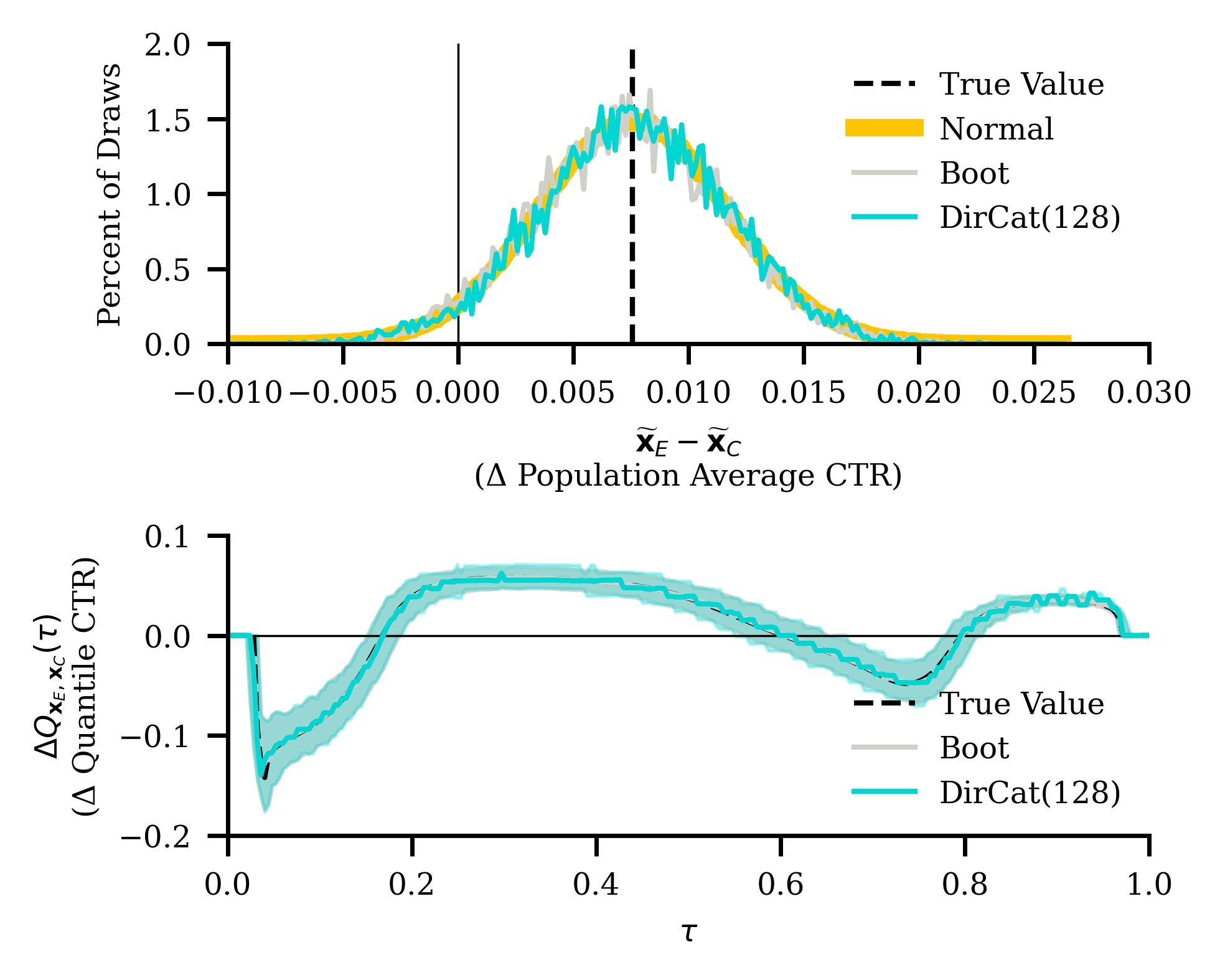}
  \else
    \includegraphics[width=10cm, height=8cm]{example-image-a}
  \fi
  \caption{An example simulation. The top shows the various methods of creating a distribution for the difference in the average click-through rate. The bottom shows two methods for creating uncertainty intervals for the quantile differences.}
  \label{fig:simulation_example}
\end{figure}

In Fig. \ref{fig:simulation_example}, we plot the results of a single simulation. In the top subplot, all three distributions for the population average difference overlap both each other and the ground truth. The bottom subplot shows a similar result for the quantiles (one bootstrap estimated, the other Dirichlet-Categorical with 128 bins estimated).

We were interested in three specific results for each simulation. Firstly, was there a bias in the Dirichlet-Categorical approach that made it an unreliable estimator of the ground truth statistical values? Secondly, are the $99\%$ credibility intervals for the mean difference and quantile difference wider than other estimation procedures? And finally, do those credibility intervals contain the ground truth values the expected percent of time?

We tackle the bias question first. To not overload terms, we use ``offset'' to describe the difference between the Riemannian-estimated ground truth for a statistic (mean difference, chance to beat, expected loss, and quantile difference) and its estimate (either via the sample-statistics-constructed Normal distribution, bootstrapping, or a Dirichlet-Categorical approximation). These offsets were calculated and stored for each of the $10,000$ simulations run. Across the simulations, the expected value of the offsets should be 0 and ideally have low variance.

\begin{figure}[!htbp]
  \centering
  \ifnum 1=\plotmyimages 
    \includegraphics[scale=0.85]{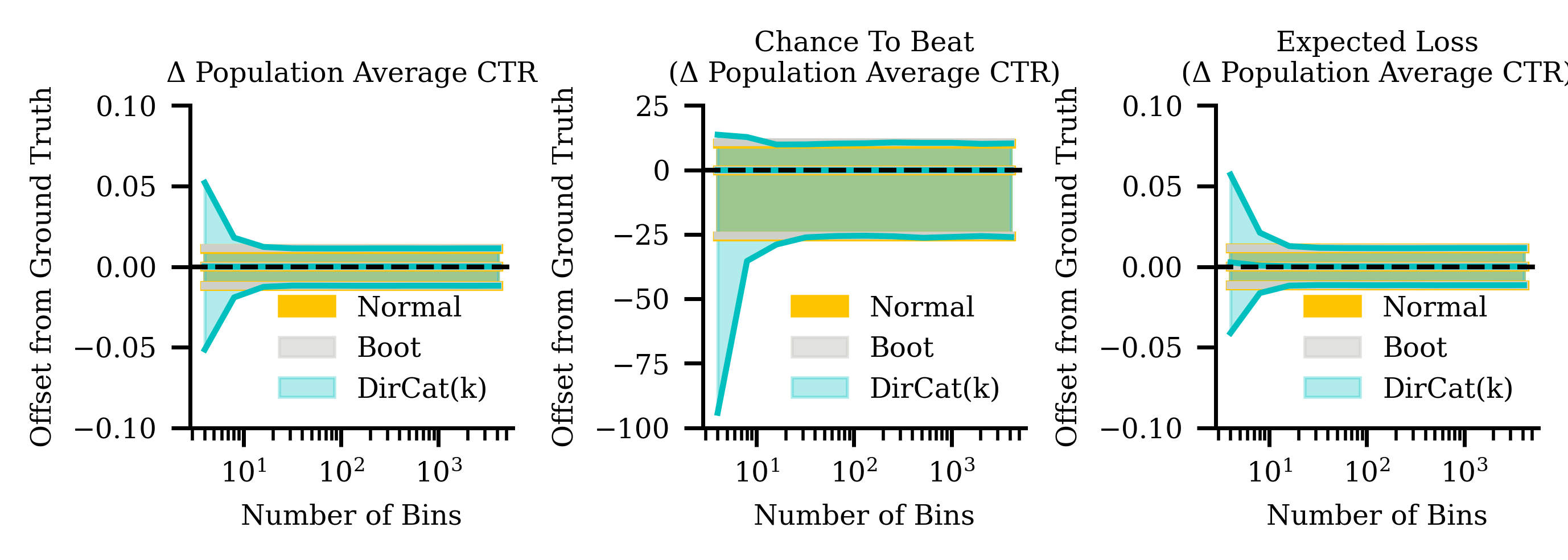}
  \else
    \includegraphics[width=12cm, height=4cm]{example-image-a}
  \fi
  \caption{The various statistics and their ability to capture the true value. The shaded error bars are the $99\%$ uncertainty intervals and the solid lines are the median values. The x-axis is the number of bins used in the Dirichlet-Categorical approximation. The Normal and bootstrap method results are extended throughout the bin range as a visual aid.}
  \label{fig:simulation_true_offset}
\end{figure}

We plot the $99\%$ uncertainty intervals for the offsets of the mean difference, chance to beat, and expected loss in Fig. \ref{fig:simulation_true_offset}. Once you've reached roughly 32 bins, all estimation procedures have a mean offset $\approx0$ and similar variance for the offset. This suggests that, with just a few dozen bins, the Dirichlet-Categorical procedure performs as well as bootstrapping and the sample-statistics-constructed Normal distribution.

\begin{figure}[!htbp]
  \centering
  \ifnum 1=\plotmyimages 
    \includegraphics[scale=0.85]{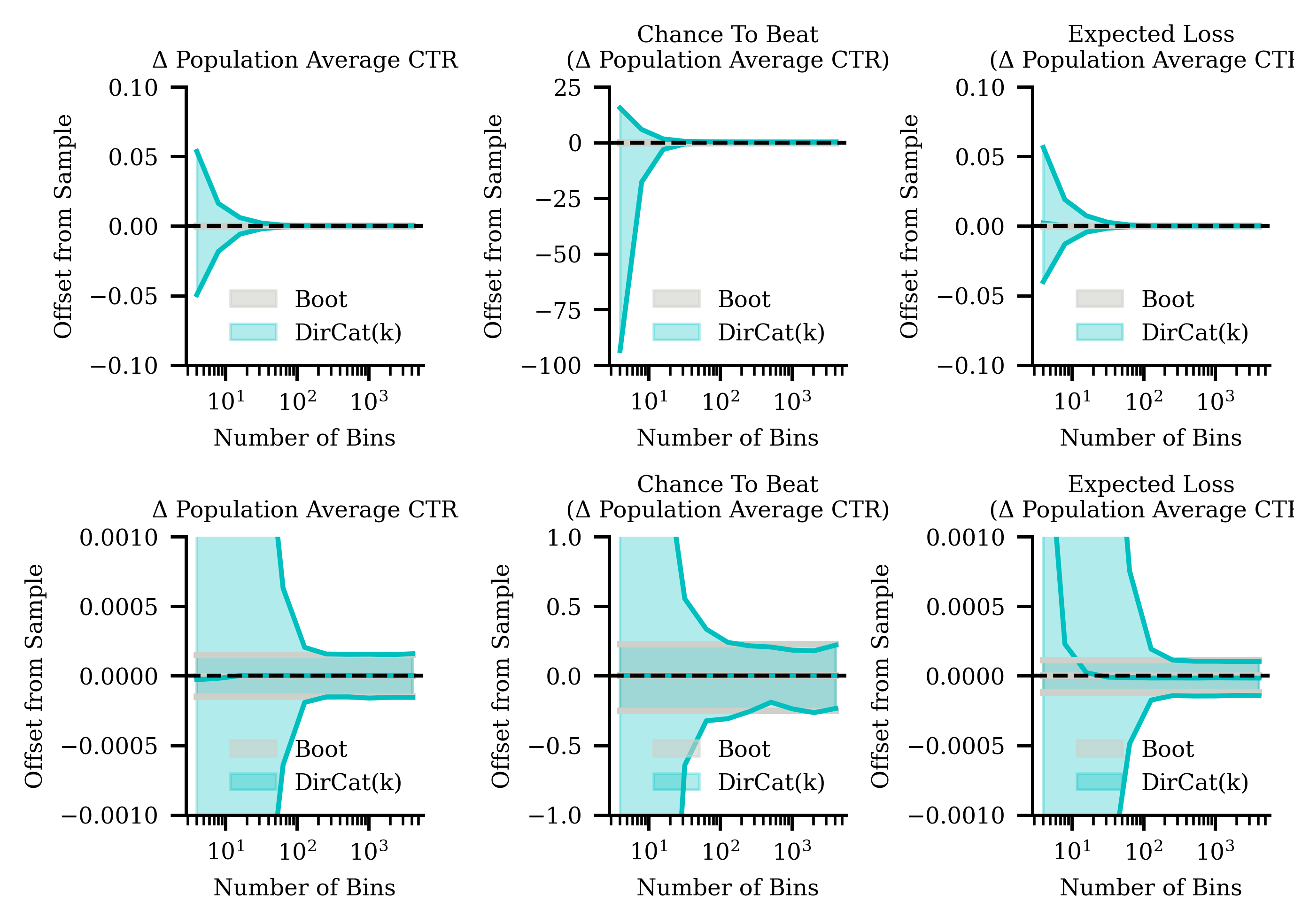}
  \else
    \includegraphics[width=12cm, height=8cm]{example-image-a}
  \fi
  \caption{(top) As in Fig. \ref{fig:simulation_true_offset}, but comparing the sample-statistics-derived Normal distribution with the bootstrap and Dirichlet-Categorical approximation. (bottom) The same data as the top, but with tighter bounds.}
  \label{fig:simulation_empirical_offset}
\end{figure}

We hypothesized that the large offset variance at small bin numbers is due to the aforementioned issue concerning non-uniformly-distributed data within the bin interval. We also hypothesized that a significant chunk of the remaining variance was due to the randomness of the sample taken from the population. This offset due to sampling should be similar in all our estimation procedures within a single simulation. Thus, we compared all the statistics from the sample-statistics-constructed Normal distribution to the bootstrap and Dirichlet-Categorical approximation in Fig. \ref{fig:simulation_empirical_offset}. The top row uses the same y-axis limits as Fig. \ref{fig:simulation_true_offset}, demonstrating the drastically-reduced offset variance at larger bin counts. The bottom row tightens the y-axis, showing that with more bins the Dirichlet-Categorical approximation is similar to bootstrapping. 

\begin{figure}[!htbp]
  \centering
  \ifnum 1=\plotmyimages 
    \includegraphics[scale=0.75]{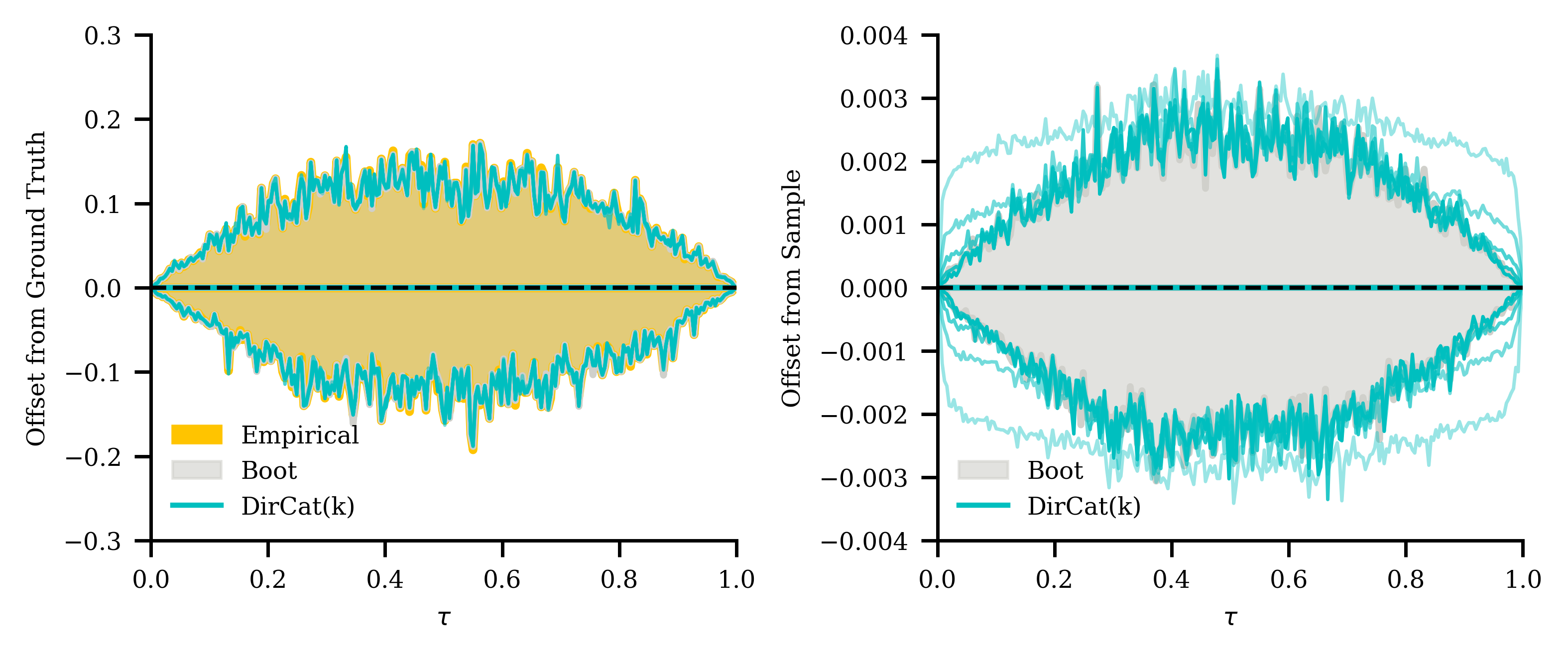}
  \else
    \includegraphics[width=12cm, height=5cm]{example-image-a}
  \fi
  \caption{(left) The various quantiles and their ability to capture the true value. The shaded error bars are the $99\%$ uncertainty intervals and the solid lines in the middle are the median values. For visualization purposes, we only plot the Dirichlet-Categorical approximation with 256, 512, 1024, 2048, and 4096 bins (using increasingly darker shades of cyan to indicate the number of bins). (right) Same as in the left subplot, but comparing the empirical CDF with the bootstrap and Dirichlet-Categorical approximation. Note the two orders of magnitude change in the y-axis limits.}
  \label{fig:simulation_both_offset_quantiles}
\end{figure}

We also plot the $99\%$ uncertainty intervals for the quantile difference offsets with ground truth and with the empirical CDF in Fig. \ref{fig:simulation_both_offset_quantiles}. As expected from a more fine-grained measure, more bins are needed to reduce the variance in the offset. Once you've reached roughly 256 bins the Dirichlet-Categorical approximation is similar to bootstrapping. The offset between the quantiles of the ground truth and any estimator is quite large, which is also expected and aligns with our practical experience. It usually takes a few days of data to tighten the uncertainty intervals around a quantile estimate. 

These results indicate that the Dirichlet-Categorical procedure (particularly with a larger number of bins) is an unbiased estimator with a similar variance in the offset as the central-limit theorem's distribution of the sample mean or a bootstrap.

\begin{figure}[!htbp]
  \centering
  \ifnum 1=\plotmyimages 
    \includegraphics[scale=1]{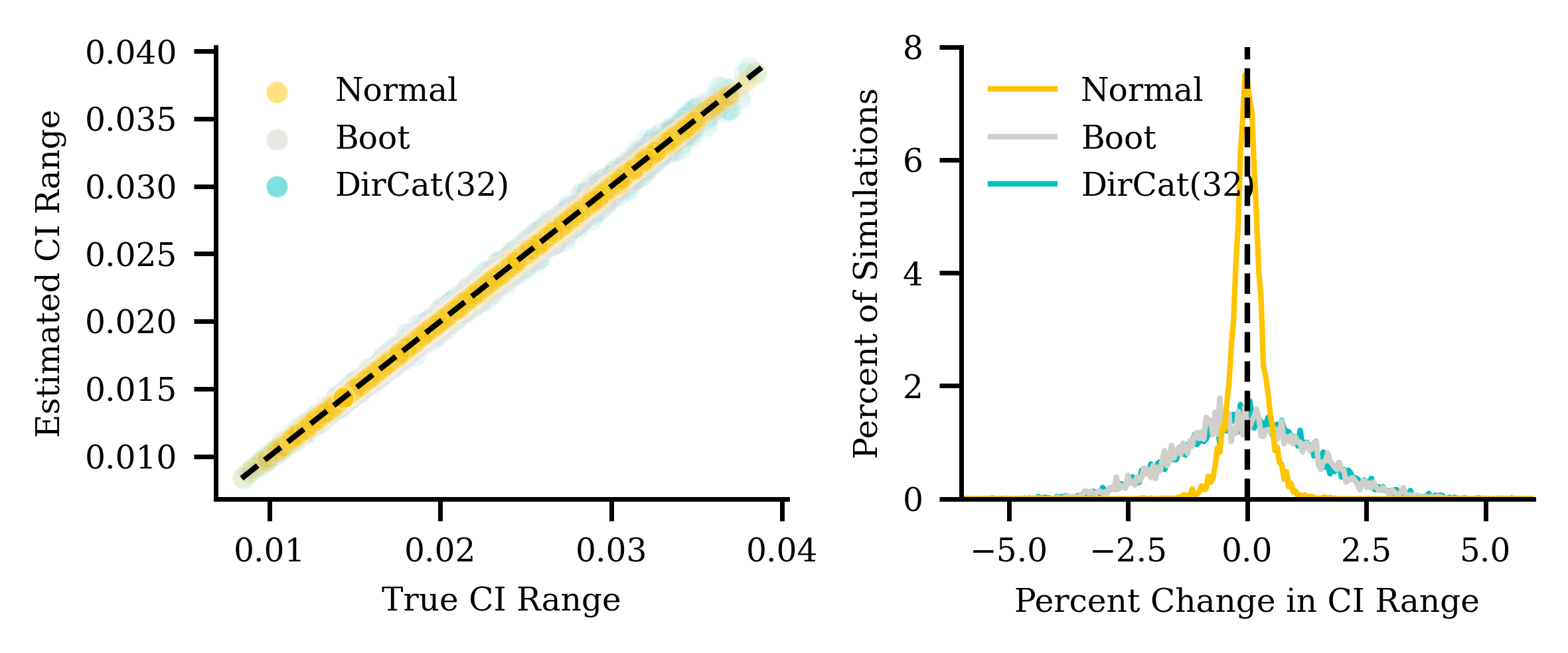}
  \else
    \includegraphics[width=12cm, height=5cm]{example-image-a}
  \fi
  \caption{(left) The various estimators $99\%$ uncertainty intervals and the true interval (using the known population parameters). The dashed line is unity. (right) The distribution of estimator ranges (as a normalized percent of the ground truth).}
  \label{fig:simulation_ci_ranges}
\end{figure}

Next, we investigated whether the range of the $99\%$ uncertainty intervals were similar between estimation procedures. We focus on the mean difference statistic since all estimation procedures and the ground truth can then be compared. Within each simulation, we calculate the $99\%$ uncertainty intervals using each estimator. For comparison, we also calculate the interval using the Normal distribution (for the sample mean) created from the true population mean, true population variance, and number of visitors. We then calculate and store the range for each interval. In Fig. \ref{fig:simulation_ci_ranges} we plot these ranges. As can be seen in the left subplot, each estimator's range lies closely along the diagonal, indicating the ranges are strongly correlated with little deviation. In the right subplot, we capture the percent change from the true confidence interval range. While all display no bias, the sample-statistics-constructed Normal distribution has tighter bounds than either the bootstrap or Dirichlet-Categorical procedures. This result suggests that, should your only concern be the posterior of the difference in population means, the central limit theorem should be invoked and a Normal model used. However, in practice, during the transition period, we did not notice experiments being underpowered or taking longer. We welcome feedback readers might have concerning the calculation of theoretical bounds for these uncertainty intervals. 

\begin{figure}[!htbp]
  \centering
  \ifnum 1=\plotmyimages 
    \includegraphics[scale=1]{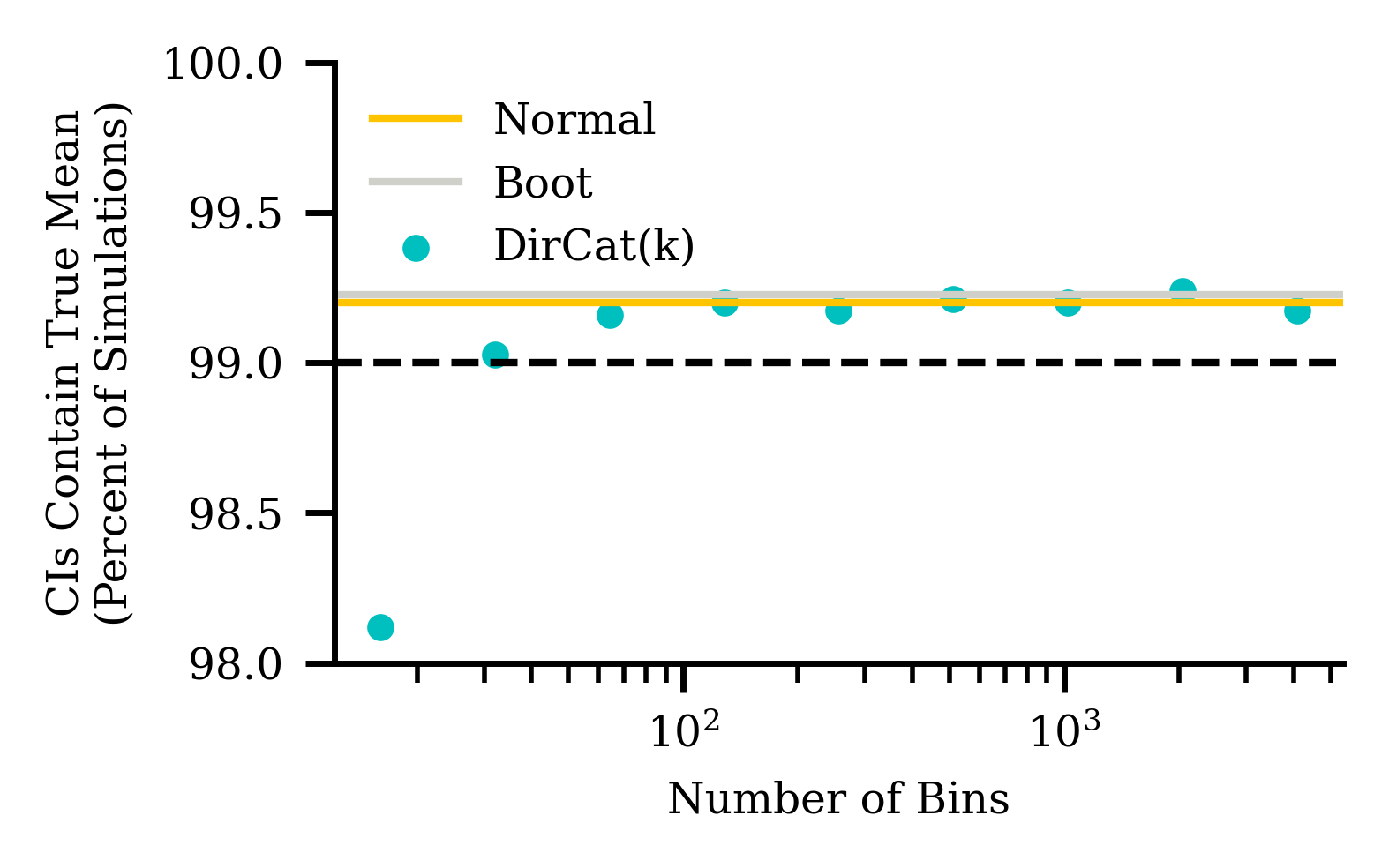}
  \else
    \includegraphics[width=12cm, height=5cm]{example-image-a}
  \fi
  \caption{The various estimators $99\%$ uncertainty intervals contain the ground truth mean difference in $\sim99\%$ of simulations (as expected).}
  \label{fig:simulation_ci_true_mean}
\end{figure}

We finally wanted to test whether the $99\%$ credibility intervals generated from the Dirichlet posterior contained the ground truth in $99\%$ of our simulations (as would be expected). As Fig. \ref{fig:simulation_ci_true_mean} shows, each estimator (beyond a few dozen bins) adheres to this principle.

In summary, we believe these results demonstrate that the Dirichlet-Categorical approximation is an unbiased estimator, has uncertainty intervals akin to bootstrapping, and does indeed contain the ground truth the expected number of times. These positive results led us to wrap the details into a package that our experimentation teams use to analyse A/B tests.

\section{Conclusion}

Altogether, the Dirichlet-Categorical framework allows for rapid, simple, flexible posterior approximations of realistic datasets in online marketplaces. It scales to massive data sets and is easily updated online via simply adding incoming data to the appropriate bin. Priors can be constructed using historical data; merely place them into the appropriate bin. The amount of historical data provided can then act as a modulator of prior strength. We have presented functions upon this posterior simplex that evaluate averages and quantiles, both of which are instrumental in understanding complex distributions. The framework can easily extend beyond these statistics and should be adaptable to the reader's own business needs. 

\section*{Acknowledgments}

We thank Peter A. Appleby for helpful revisions, Josh Rickards for thoughtful criticisms, and the numerous data scientists at Autotrader for their ongoing feedback. We further thank Pantelis Hadjipantelis for his insightful review and suggestions that greatly improved the final paper.

\bibliographystyle{plainnat}
\bibliography{references}

\end{document}